\documentclass[12pt]{article}
\usepackage{amsxtra,amssymb,amsthm,amsmath,latexsym}

\theoremstyle{plain}
\newtheorem{definition}{Definition}[section]
\newtheorem{theorem}{Theorem}[section]
\newtheorem{lemma}{Lemma}[section]

\newtheorem{remark}{Remark}[section]
\numberwithin{equation}{section}

\def\nd{\noindent}

\def\R{{\mathbb R}}

\def\C{{\mathbb C}}
\def\oH{\buildrel\circ\over H}
\def\oH1{\buildrel\circ\over H\kern-.02in{}^1}
\def\qed{{\hfill $\Box$}}
\def\l{\ell}

\def\supp{\hbox{\,supp\,}}
\def\const{\hbox{\,const\,}}
\def\arctg{\hbox{\,arctg\,}}

\begin{document}

\begin{titlepage}

\title{Stability of the solutions to 3D inverse scattering problems with
fixed-energy data.
   \thanks{key words: inverse scattering, stability estimates}
   \thanks{Math subject classification:35R30, 35R25, 47H17, 65M30,
   65L20, 34C35, 34G20}
}

\author{
A.G. Ramm\\
 Mathematics Department, Kansas State University, \\
 Manhattan, KS 66506-2602, USA\\
ramm@math.ksu.edu\\
}

\date{}

\maketitle\thispagestyle{empty}

\begin{abstract}
A review of the author's results is given. Inversion formulas 
and stability
results for the solutions to 3D inverse scattering problems with
fixed energy data are obtained. Inversion of exact and noisy data is
considered. The inverse potential scattering problem with
fixed-energy scattering data is discussed in detail, inversion
formulas for the exact and for noisy data are derived, error
estimates for the inversion formulas are obtained.
The inverse obstacle scattering problem is considered
for non-smooth obstacles. Stability estimates are derived for
inverse obstacle scattering problem in the class of smooth
obstacles. Global estimates for the scttering amplitude
are given when the potential grows to infinity in a bounded
domain. Inverse geophysical scattering problem is discussed
briefly. An algorithm for constructing the Dirichlet-to-Neumann
map from the scattering amplitude and vice versa is 
obtained.
An analytical example of non-uniqueness of the solution to a 
3D inverse problem of geophysics and a uniqueness theorem
for an inverse problem for parabolic equations are given.
 \end{abstract}

\end{titlepage}

\tableofcontents

\newpage

%%%%%%%%%%%%%%%%%%%%%%%%%%%%%%%%%%%%%%%%%
\section{Introduction }%1
In this paper 3D inversion scattering problems with fixed-energy data
are discussed. These problems include inverse problems of potential,
obstacle, and geophysical scattering (IPS, IOS, IGS).

Inverse potential scattering problem is discussed in detail: uniqueness of
its solution, reconstruction formulas for inversion of the exact data and for
inversion of noisy data are given and error estimates for these formulas are
obtained. These estimates yield the stability estimates for the solution of
the inverse scattering problem.

For the inverse obstacle scattering the uniqueness theorem is proved for rough
domains, stability estimates are obtained for $C^{2, \lambda}$ domains,
$0 < \lambda < 1$, that is, for domains whose boundary in local coordinates
is a graph of $C^{2, \lambda}$ function. Reconstruction formulas are
discussed.

For inverse geophysical scattering the inverse scattering problem is reduced
to inverse scattering problem for a potential.

Construction of the Dirichlet-to -Neumann map from the scattering data and
vice versa is given. Analytical example of nonuniqueness of the solution
of an inverse 3D problem of geophysics is given.

The results discussed in this paper were obtained
mostly by the author, see \cite{R10}, \cite{R1}-\cite{R45},
however
the presentation and some of the estimates are improved
in this paper. Only some selected results from the cited papers
are included in this review.

%%%%%%%%%%%%%%%%%%%%%%%%%%%%%%%%%%%%%%%%%
\subsection{The direct potential scattering problem.}
We want to study the inverse potential scattering problem of finding $q(x)$
given some scattering data.

Consider the direct scattering problem first and let us formulate some
basic results which we need.

Let
\begin{equation}
        [\nabla^2 + k^2 - q(x)] u(x, \alpha, k) = 0
        \quad \hbox{\ in\ } \R^3, \quad
        x \in \R^3,
        \end{equation}
\begin{equation}
        u = e^{ik \alpha \cdot x} + A(\alpha^\prime, \alpha, k)
        \frac{e^{ikr}}{r} + o \left(\frac{1}{r}\right), \quad
        r := |x| \to \infty, \quad
        \alpha^\prime := \frac{x}{r}
        \end{equation}

Here $u(x,k)$ is the scattering solution, $k = \const >0$ is fixed. Without
loss of generality we take $k=1$ in what follows unless other choice is
suggested explicitly. A unit vector $\alpha \in S^2$ is given, where $S^2$ is
the unit sphere in $\R^3$. Vector $\alpha$ has a physical meaning of the
direction of the
incident plane wave, while $\alpha^\prime \in S^2$ is the direction
of the scattered wave, $k^2$ is the fixed energy. The function
$A(\alpha^\prime, \alpha, k)$ is called the scattering amplitude. It
describes the first term of the asymptotics of the scattered field as
$r \to \infty$ along the direction $\alpha^\prime = \frac{x}{r}$.

The function $q(x)$ is called the potential. We assume that
\begin{equation}
  \begin{split}
     q \in Q
        & := Q_a \cap L^\infty (\R^3),      \\
      Q_a
        & := \{ q: q(x) = \overline{q(x)}, \quad
        q(x) \in L^2(B_a), \quad q(x) = 0
        \hbox{\ if\ } |x|>a \},
        \end{split}
        \end{equation}
where $a >0$ is an arbitrary large fixed number which we call the range of
$q(x)$, and the overbar stands for complex conjugate.

In many results $q \in Q_a$ is sufficient, but $q \in Q$ is used in the
proof of a crucical estimate (2.17) below.

%%%%%%%%%%%%%%%%%%%%%%%%%%%%%%%%%%%%%%%%%
\subsection{Review of the known results.}
Let us formulate some of the known results about the solution to problem
(1.1)-(1.2), the scattering solution. These results can be found in many
books, for example, in the appendix to \cite{R1}, where a brief but
self-contained presentation of the scattering theory is given.

\subsubsection{The scattering problem has a unique solution if
$q \in Q_a$.}
In fact, the above
result is proved for much larger class of $q$ (\cite{P}, \cite{H}), but for
inverse scattering problem with noisy data it is necessay to assume $q(x)$
compactly supported \cite{R2}. 
Indeed,  represent  the potential
$q(x)$ as $q=q_1+q_2$, where $q_1=0$ for $|x|>a$ and $q_1=q$ for $|x|\leq a$.
Call $q_2$ the tail of the potential $q$.
If one assumes a priori that $q=O(|x|^{-b}),$ where $b>3$, then the contribution 
of the tail of the potential to the scattering amplitude is 
of order $O(|a|^{3-b})$ and tends to $0$ as $a\to \infty.$
At some value of $a,$ say at $a=a_0,$
 this contribution becomes of the order of the noise in the scattering data.
One cannot, in principle, discriminate between the noise and the contribution of the tail
of the potential for $a>a_0$. 
Therefore the tail of $q$ for $a>a_0$ cannot be determined from noisy data.

One has
\begin{equation}
        \sup_{x \in \R^3} |u(x,k)| \leq c, \quad k = \const >0.
        \end{equation}
By $c>0$ we denote various constants. If $q \in Q_a$ then $u(x,k)$ extends
as a meromorphic function to the whole complex $k$-plane. Let $G(x,y,k)$ denote the
resolvent kernel of the self-adjoint Schr\"odinger operator
$Lu = - \nabla^2 + q(x)$ in $L^2(R^3)$:
\begin{equation}
        (L-k^2) G(x,y,k) = \delta (x-y) \hbox{\ in\ } \R^3,
        \end{equation}
\begin{equation}
        \lim_{r \to \infty} \int_{|x| = r} \left|
        \frac{\partial G}{\partial |x|} - ik G \right|^2 ds = 0, \quad
        y \hbox{\ is\ } \hbox{\ fixed\ }, \quad k>0.
        \end{equation}

The function $u(x,k)$ can be defined by the formula:
\begin{equation}
        G(x,y,k) = \frac{e^{ik|y|}}{4 \pi |y|} u(x, \alpha, k) +
        o \left(\frac{1}{|y|} \right), \quad
        \frac{y}{|y|} = - \alpha,
        \end{equation}
where
$o \left(\frac{1}{|y|}\right) = O\left(\frac{1}{|y|^2}\right)$
is uniform with respect to $x$ varying in compact sets and formula (1.7)
can be differentiated with respect to $x$ \cite{R2}, \cite{R7}.

The function $G(x,y,k)$ is a meromorphic function of $k$ on the whole complex
$k$-plane. It has at most finitely many simple poles
$ik_j$, $k_j >0$, $1 \leq j \leq J$ in $\C_+ := \{k: Im k>0 \}$
and if $q(x) \not\equiv 0$, $q \in Q_a$, infinitely many poles, possibly not
simple, in $\C_- = \C \setminus \overline{\C_+}$. There are no poles on the
real line except, possibly at $k=0$.

The functions $u_j(x) \in L^2 (\R^3)$, solving (1.1) with $k=ik_j$, are
called eigenfunctions of the discrete spectrum of $L$, $-k^2_j$ are the
negative eigenvalues of $L$. There are at most finitely many of these if
$q \in Q_a$.

The eigenfunction expansion formulas are known:
$$f(x) = \sum_j f_j u_j(x) + \int_{\R^3} \widetilde{f}
(\xi) u(x, \xi) d \xi, \quad
  |\xi| = k, \quad \xi=k \alpha,$$
where
$$ f_j := (f, u_j)_{L^2(\R^3)}, \quad \widetilde{f}(\xi) =
\frac{1}{(2 \pi)^3}
\int_{\R^3} f(x) \overline{u(x, \alpha, k)} dx,$$
(see e.g. \cite{R1}).

If $E_\lambda$ is the resolution of the identity of the selfadjoint operator
$L$, and $E_\lambda(x,y)$ is its kernel, then
$$ \frac{d E_\lambda (x,y)}{d \lambda} = \frac{1}{\pi} Im
   G(x,y, \sqrt{\lambda}) = \frac{\sqrt{\lambda}}{16 \pi^3}
   \int_{S^2} u(x, \alpha, \sqrt{\lambda})
   \overline{u(y, \alpha, \sqrt{\lambda})} d \alpha, \quad \lambda >0.$$

\subsubsection{Properties of the scattering amplitude}

The scattering amplitude has the following well-known properties (see e.g.
\cite{R1}):
\begin{equation}
  \begin{split}
  A(\alpha^\prime, \alpha, k)
     &= A(-\alpha, -\alpha^\prime, k)  \quad \hbox{(reciprocity)},
     \notag \\
  \overline{A(\alpha^\prime, \alpha, k)}
     &=A(\alpha^\prime, \alpha, -k), \quad k>0 \quad \hbox{(reality)},
     \notag \\
  Im A(\alpha^\prime, \alpha, k)
     &= \frac{k}{4 \pi} \int_{S^2} A(\alpha^\prime, \beta,k)
        \overline{A(\alpha, \beta, k)}
        d\beta, \quad k>0 \quad \hbox{(unitarity)}.
        \notag
  \end{split}
  \end{equation}

In particular,
$$Im A(\alpha, \alpha, k) = \frac{k}{4 \pi} \int_{S^2}
 |A(\alpha, \beta, k)|^2 d \beta \quad \hbox{(optical theorem)}.$$
If $q \in Q_a$, and $k=1$, then the scattering amplitude is an
analytic function
of $\alpha^\prime$ and $\alpha$ on the algebraic variety
\begin{equation}
        M := \{ \theta : \theta \in \C^3, \theta \cdot \theta =1\}, \quad
        \theta \cdot w := \sum^3_{j=1} \theta_j w_j.
        \end{equation}
This variety is non-compact, intersects $\R^3$ over $S^2$, and, given any
$\xi \in \R^3$, there exist (many) $\theta, \theta^\prime \in M$ such that
\begin{equation}
        \theta^\prime - \theta = \xi, \quad |\theta| \to \infty, \quad
        \theta, \theta^\prime \in M.
        \end{equation}

In particular, if one chooses the coordinate system in which
$\xi = te_3$, $t>0$, $e_3$ is the unit vector along the $x_3$-axis, then
the vectors
\begin{equation}
        \theta^\prime = \frac{t}{2}e_3 + \zeta_2e_2 + \zeta_1e_1, \quad
        \theta = -\frac{t}{2}e_3 + \zeta_2e_2+\zeta_1 e_1, \quad
        \zeta^2_1 + \zeta^2_2 = 1-\frac{t^2}{4},
        \end{equation}
satisfy (1.9) for any complex numbers $\zeta_1$ and $\zeta_2$ satisfying the last
equation in (1.10) and such that $|\zeta_1|^2+|\zeta_2|^2 \to \infty$. There are
infinitely many such $\zeta_1, \zeta_2 \in \C$. If $q \in Q_a$ than the function
$A(\alpha^\prime, \alpha, k)$ is a meromorphic function of
$k \in \C$ which has poles at the same points as $G(x,y,k)$.

One has
\begin{equation}
        -4 \pi A(\alpha^\prime, \alpha, k) = \int_{B_a}
        e^{-ik \alpha^\prime \cdot x} q(x) u(x, \alpha, k) dx.
        \end{equation}
The $S$-matrix is defined by the formula
\begin{equation}
        S=I + \frac{ik}{2 \pi} A, \quad S^\ast S = SS^\ast = I,
        \end{equation}
and is a unitary operator in $L^2(S^2)$.

If $\nabla q \in Q_a$ then
\begin{equation}
        \phi (x, \alpha, k) := e^{-ik \alpha \cdot x} u(x,\alpha, k) =
        1 + \frac{1}{2ik} \int^\infty_0 q(x-r\alpha) dr +
        o\left(\frac{1}{k}\right), \quad k \to \infty.
        \end{equation}
Therefore
\begin{equation}
        q(x) = \alpha \cdot \nabla_x \lim_{k \to \infty}
        \left\{ 2ik [\phi(x,\alpha,k)-1]\right\},
        \end{equation}
and
\begin{equation}
        A(\alpha, \alpha, k) = -\frac{1}{4 \pi} \int_{\R^3} q(x)dx +
        o(1), \quad k \to +\infty.
        \end{equation}

\subsubsection{The fundamental equation.}  
Denote $u^+ := u(x, \alpha,k), u^- := u(x, -\alpha, -k)$, $k>0$. Then
$u^+=Su^-$, that is
\begin{equation}
        u^+ = u^- + \frac{ik}{2 \pi} \int_{S^2}
        A(\alpha^\prime, \alpha, k) u^-(x, \alpha^\prime, k) d \alpha^\prime.
        \end{equation}

\subsubsection{Completeness properties of
the scattering solutions.} 

\begin{description}

\item[\quad a)]
{\it  If $h(\alpha) \in L^2(S^2)$ and
\begin{equation}
        \int_{S^2} h(\alpha) u(x, \alpha, k)d \alpha = 0 \quad
        \forall x \in B^\prime_R := \{x:|x|>R\}, \quad
        k>0 \hbox{\ is\ } \hbox{\ fixed\ }
        \end{equation}
then $h(\alpha)=0$.
}

Let $N_D(L) = \{w: Lw=0$ in $D$, $w \in H^2(D)\}$,
where $D \subset \R^3$ is a bounded domain, $H^2(D)$ is the Sobolev space.

\item[\quad b)]
The set $\left\{u(x,\alpha, k)\right\}_{\forall \alpha \in S^2}$
is total in $N_D(L-k^2)$, that is, for any
$\varepsilon >0$, however small,  and any fixed
$w \in N_D (L-k^2)$, there exists $\nu_\varepsilon (\alpha) \in L^2(S^2)$
such that
\begin{equation}
        \Vert w(x) - \int_{S^2} u(x,\alpha, k) \nu_\varepsilon
        (\alpha) d \alpha \Vert_{H^2(D)} < \varepsilon.
        \end{equation}

The $\nu_\varepsilon (\alpha)$ depends on $w(x)$.
\end{description}

\subsubsection{Special solutions} 
There exists $\psi(x, \theta, k)$, $\psi \in N_D (L-k^2)$, such that
\begin{equation}
        [\nabla^2 + k^2 - q(x)] \psi = 0 \hbox{\ in\ }
        \R^3, \quad \psi = e^{ik \theta \cdot x} [1 + R(x, \theta, k)],
        \quad\theta \in M,
        \end{equation}
\begin{equation}
        \Vert R \Vert_{L^\infty(D)} \leq c
        \frac{(\ln |\theta|)^{\frac{1}{2}}}{|\theta|^{\frac{1}{2}}}, \quad
        |\theta| \to \infty, \quad \theta \in M,
        \end{equation}
\begin{equation}
        \Vert R \Vert_{L^2(D)} \leq \frac{c}{|\theta|}, \quad
        |\theta| \to \infty, \quad \theta \in M,
        \end{equation}
where $D\subset \R^3$ is an arbitrary bounded domain.

\subsubsection{Property $C$ for the pair $\{L_1-k^2, L_2-k^2\}$} 

Let
$L_jw := \sum^{M_j}_{|m|=0} a_{jm}(x) \partial^{|m|} w(x)$, $x \in \R^n$,
$n \geq 2$, $j=1,2,$
be linear formal partial differential operators, that is, formal differential
expressions.

Let
$N_j = N_{jD}(L_j) := \{w : L_j w=0 \hbox{\ in\ } D \}$,
where $D \subset R^n$ is an arbitrary fixed bounded domain and the equation
is understood in the sense of the distribution theory. Consider the subsets
of $N_j$, $j=1,2,$ which form an algebra in the sense that the products
$w_1 w_2 \in L^{p^\prime}(D)$, where $w_j \in N_j$,
$p^\prime = \frac{p}{p-1}$, and $1 \leq p \leq \infty$. If $p=1$ define
$p^\prime = \infty$, and if $p= \infty$ define $p^\prime =1$. We write
$\forall w_j \in N_j$ meaning that $w_j$ run through the above subsets of
$N_j$.

\begin{definition} 
We say that the pair of linear partial differential operators
$\{L_1, L_2\}$ has property $C_p$ if and only if the set
$\{w_1 w_2\}$ is total in $L^p(D)$, that is, if $f(x) \in L^p(D)$ and

\begin{equation}
        \int_D f(x) w_1(x) w_2(x) dx = 0 \quad \forall w_j \in N_j, j=1,2,
        \end{equation}
then
\begin{equation}
    f(x) =0.
    \end{equation}
If the above holds for any $p\geq 1,$ we say that property $C$ holds
for the pair $\{L_1, L_2\}$.
\end{definition}

\begin{theorem}{1.1 \cite{R2}.}
Let $L_j = -\nabla^2 + q_j (v)$, $q_j(x) \in Q_a$, $k = \const \geq 0$
is arbitrary fixed. Then the pair
$\{L_1 -k^2, L_2 -k^2 \}$ has property $C$.
\end{theorem}

\begin{proof}
Note that $\psi_j \in N_j$, $j =1,2$, where $\psi_j$ are defined in section
1.2.5 above. Without loss of generality take $k=1$, let
$\psi (x, \theta, 1) := \psi(x, \theta)$.
One has
$$\psi_1 (x, \theta^\prime) \psi_2 (x, -\theta) =
e^{i(\theta^\prime - \theta) \cdot x} (1 +R_1) (1 + R_2).$$
Choose $\theta^\prime, \theta \in M$ such that (1.9) holds with an arbitrary
fixed $\xi \in \R^3$. Then
\begin{equation}
        \psi_1 (x, \theta^\prime) \psi_2 (x, -\theta) = e^{i \xi \cdot x}
        (1 + o(1)) \hbox{\ as\ } |\theta| \to \infty.
        \end{equation}
Since the set $\{e^{i \xi \cdot x}\}_{\forall \xi \in \R^3}$
is total in $L^p(D)$, $p \geq 1$, $D \subset \R^n$ is a bounded domain,
the conclusion of Theorem 1.1 follows.
\end{proof}

\begin{remark}
One cannot take unbounded domain $D$ in the above argument because $o(1)$ in
(1.24) holds for bounded domains.
\end{remark}

One can take the space of $f(x)$ larger than $L^1(D)$, for example, the space
of distribution of finite order of singulartiy if $q(x)$ is sufficiently
smooth \cite{R2}.   %\cite{GR}.

\begin{theorem}
The set
$\{u_1(x, \alpha, k) u_2 (x, \beta, k)\}_{\forall \alpha, \beta
\in S^2, k >0}$ is complete in $L^p(D)$,
where $D \subset \R^3$ is an arbitrary fixed bounded domain, and  $p \geq 1$
is fixed.
\end{theorem}

\begin{proof}
The conclusion of Theorem 1.2 follows from Theorem 1.1 and (1.18).
\end{proof}

\subsubsection{Properties of the Fourier coefficients of
$A(\alpha^\prime, \alpha)$.} 

We denote $A(\alpha^\prime, \alpha, k)|_{k=1}:= A(\alpha^\prime, \alpha)$,
and write
\begin{equation}
        A(\alpha^\prime, \alpha) = \sum^\infty_{\l =0} A_\l (\alpha)
        Y_\l (\alpha^\prime), \quad A_\l (\alpha) := \int_{S^2}
        A(\alpha^\prime, \alpha) \overline{Y_l(\alpha^\prime)}
        d \alpha^\prime,
        \end{equation}
where $Y_\l (\alpha^\prime) = Y_{\l,m} (\alpha^\prime), -\l \leq m \leq \l$,
summation over $m$, is understood in (1.25) and similar formulas below,
e.g. (1.31), (1.37), etc,
\begin{equation}
        \begin{split}
        &Y_{\l,m}(\alpha) = \frac{(-1)^mi^\l}{\sqrt{4 \pi}}
        \left[\frac{(2\l+1)(\l- m )!}{(\l +  m )!} \right]^{\frac{1}{2}}
        e^{im \varphi} P_{\l, m} (\cos \vartheta), \\
        &\overline{Y_{\l,m}(\alpha)} = (-1)^{\l +m} Y_{\l,-m}
(\alpha),\,\,\,
        Y_{\l,m} (-\alpha)= (-1)^\l Y_{\l,m} (\alpha).
        \end{split}
        \end{equation}
Here $P_{\l,m } (\cos \vartheta) = (\sin \vartheta)^{ m }
\frac{d^{ m } P_\l (\cos \vartheta)}{(d \cos \vartheta)^m}$,
$0\leq m\leq\l$, $P_\l(x)$
is the Legendre polynomial, $(\vartheta, \varphi)$ are the angles
corresponding to the point $\alpha \in S^2,$
$P_{\l,-m}(\cos\vartheta)
  =(-1)^m\frac{(\l-m)!}{(\l+m)!} P_{\l,m}(\cos\vartheta)$,
$0\leq m\leq\l$.

Consider a subset $M^\prime \subset M$ consisting of the vectors
$\theta = (\sin \vartheta \cos \varphi, \sin \vartheta \sin \varphi, \cos \vartheta)$
where $\vartheta$ and $\varphi$ run through the whole complex plane. Clearly
$\theta \in M$, but $M^\prime$ is a proper subset of $M$. Indeed, any
$\theta \in M$ with $\theta_3 \neq \pm 1$ is an element of $M^\prime$.
If $\theta_3 = \pm 1$, then $\cos \vartheta = \pm 1$, so
$\sin \vartheta = 0$ and one gets $\theta = (0,0, \pm 1) \in M^\prime$. However,
there are vectors $\theta = (\theta_1, \theta_2, 1) \in M$ which do not
belong to $M^\prime$. Such vectors one obtains choosing
$\theta_1, \theta_2 \in \C$ such that $\theta^2_1 + \theta_2^2 = 0$.
There are infinitely many such vectors. The same is true for vectors
$(\theta_1, \theta_2, -1)$. Note that in (1.9) one can replace $M$ by
$M^\prime$ for any $ \xi \in \R^3$, $\xi \neq 2e_3$.

Let us state two estimates (\cite{R6}):
\begin{equation}
        \sup_{\alpha \in S^2} \left| A_\l (\alpha) \right| \leq
        c \left(\frac{a}{\l}\right)^{\frac{1}{2}}
        \left(\frac{ae}{2\l}\right)^{\l +1},
        \end{equation}
and
\begin{equation}
        \left|Y_\l (\theta)\right| \leq \frac{1}{\sqrt{4 \pi}}
        \frac{e^{r |Im \theta|}}{|j_\l (r)|}, \quad
        \forall r > 0, \quad \theta \in M^\prime,
        \end{equation}
where
\begin{equation}
        j_\l (r)  :=\left(\frac{\pi}{2r} \right)^{\frac{1}{2}}
        J_{\l + \frac{1}{2}} (r) = \frac{1}{2\sqrt{2}} \frac{1}{\l}
        \left(\frac{er}{2\l}\right)^\l [1 + o(1)] \hbox{\ as\ }
        \l \to \infty,
        \end{equation}
and $J_\l(r)$ is the Bessel function regular at $r=0$.
Note that $Y_\l (\alpha^\prime)$,
defined by (1.26), admits a natural analytic continuation from $S^2$ to $M$
by taking $\vartheta$ and $\varphi$ in (1.26) to be arbitrary complex numbers.
The resulting $\theta^\prime \in M^\prime \subset M$.

\subsubsection{A global perturbation formula.}  
Let $A_j(\alpha^\prime, \alpha)$ be the scattering amplitude corresponding
to $q_j \in Q_a$, $j = 1,2$. Define $A:=A_1-A_2$, $p := q_1(x) - q_2(x)$.
Then \cite{R2}
\begin{equation}
        -4\pi A(\alpha^\prime, \alpha) = \int_{B_a} p(x) u_1(x, \alpha)
        u_2(x, -\alpha^\prime) dx.
        \end{equation}

\subsubsection{Formula for the scattering solution
outside the support of the potential.}   

Let $\supp (q) \subset B_a$. The fixed-energy scattering data
$A(\alpha^\prime, \alpha) \forall \alpha^\prime, \alpha$, or, equivalently,
the data $\{A_\l (\alpha)\}_{\l = 0,1,2, \dots} \forall \alpha \in S^2$,
allow one to write an analytic formula for the scattering solution
$u(x, \alpha)$ in the region $B_a^\prime := \R^3 \backslash B_a$:
\begin{equation}
        u(x, \alpha) = e^{i \alpha \cdot x} + \sum^\infty_{\l=0}
        A_\l (\alpha) Y_\l (\alpha^\prime) h_\l (r), \quad
        r:= |x| > a, \quad \alpha^\prime := \frac{x}{r},
        \end{equation}
where $A_\l (\alpha)$ are defined in (1.25), $Y_\l(\alpha^\prime)$
are defined in (1.26),
$$h_\l(r) := e^{i \frac{\pi}{2} (\l+1)} \sqrt{\frac{\pi}{2r}}
H^{(1)}_{\l + \frac{1}{2}} (r),$$
$H^{(1)}_\l (r)$ is the Hankel function, and the normalizing factor is
chosen so that
\begin{equation}
        h_\l(r) = \frac{e^{ir}}{r} [1 + o(1)] \hbox{\ as\ } r \to \infty.
        \end{equation}

Note that [2, formula (7.1463]:
\begin{equation}
        |H^{(1)}_\l (r)|^2 = \frac{4}{\pi^2} \int^\infty_0 K_0(2r sht)
        (e^{2 \l t} + e^{-2\l t}) dt,
        \end{equation}
where $sh t := \frac{e^t-e^{-t}}{2}$.
This  formula implies that $|h_\l(r)|$ is a monotonically increasing
function of $\l$.

It is known [\cite{GR}, formula 8.478] that $r|h_\l(r)|^2$ is a monotonically
decreasing function of $r$ if $\l >0$, and
\begin{equation}
        h_\l(r) = -\frac{i^\l \sqrt{2}}{r} \left(\frac{2 \l}{er}\right)^\l
        \left[1+o(1)\right], \quad \l \to +\infty, \quad r>0.
        \end{equation}

The following known estimate can be useful:
\begin{equation}
        |Y_{\l,m} (\alpha)| \leq c \l^{\frac{m}{2}-1}, \alpha \in S^2,
        \end{equation}
where $Y_{\l,m}(\alpha)$ are the normalized in $L^2(S^2)$ spherical harmonics
(1.26).

Let us give a formula for the Green function $G(x,y, k)$ (see (1.5), (1.6))
in the region $|x|>a$, $|y| >a$, where $\supp q(x) \subset B_a$.
Let $g(x,y,k):= \frac{e^{ik|x-y|}}{4\pi |x-y|}$ and denote by
$A_{\l^\prime \l}$ the Fourier coefficients of the scattering amplitude:
\begin{equation}
        A(\alpha^\prime, \alpha) = \sum^\infty_{\l =0} A_\l(\alpha)
        Y_\l(\alpha^\prime) = \sum^\infty_{\l^\prime, \l =0}
        A_{\l^\prime \l} Y_{\l^\prime} (\alpha)
        Y_\l(\alpha^\prime).
        \end{equation}
Then
\begin{equation}
%        \begin{split}
        G(x,y,k) = 
%            &
            g(x,y,k) + \frac{k^2}{4\pi}
        \sum^\infty_{\l^\prime, \l =0}
%        \\
        A_{\l^\prime \l}
%            &
        Y_{\l^\prime} (-\alpha) Y_\l (\alpha^\prime)
        h_\l (k|x|) h_{\l^\prime} (k|y|),
        \quad |x|>a,   \quad  |y| >a,
%        \end{split}
        \end{equation}
where $\alpha^\prime := \frac{x}{|x|}$, $\alpha := \frac{y}{|y|}$.

Indeed, clearly the function (1.37) solves (1.5) in the region $|x|>a$, $|y|>a$,
where $q(x) = 0$, it satisfies (1.6), and
\begin{equation}
        \begin{split}
        G(x,y,k) =
            &\frac{e^{ik|y|}}{4\pi |y|}
        \left[e^{ik \alpha \cdot x}+ k^2 \sum^\infty_{\l^\prime, \l =0}
        A_{\l^\prime \l} Y_{\l^\prime} (\alpha) Y_\l(\alpha^\prime)
        h_\l(k|x|) \right] +o\left(\frac{1}{|y|}\right), \\
             &\hbox{\ as\ } |y| \to \infty, \ \frac{y}{|y|} = -\alpha.
        \end{split}
        \end{equation}

By (1.36), (1.31),(1.25), and (1.7), it follows that the function (1.37) has
the same main term of asymptotics (1.38) as the Green function of the
Schr\"odinger operator. Therefore the function (1.37) is identical to the
Green function (1.5)-(1.6) in the region $|x|>a$, $|y|>a$.

%%%%%%%%%%%%%%%%%%%%%%%%%%%%%%%%%%%%%%%%%
\section{Inverse potential scattering problem with fixed-energy data}%2
The IPS problem can now be formulated:
{\it given $A(\alpha^\prime, \alpha)$ $ \forall \alpha^\prime, \alpha \in S^2$,
find $q(x) \in Q_a$}. Throughout this section $k=1$.

%%%%%%%%%%%%%%%%%%%%%%%%%%%%%%%%%%%%%%%%%
\subsection{Uniqueness theorem.}   
The first result is the uniqueness theorem of Ramm \cite{R4}, \cite{R5}.

\begin{theorem}
If $q_1, q_2 \in Q_a$ and $A_1(\alpha^\prime, \alpha) =
A_2(\alpha^\prime, \alpha)$ $\forall \alpha^\prime \in S^2_1$,
$\forall \alpha \in S^2_2$, where $S^2_j$ $j=1,2,$
are arbitrary small open subsets of
$S^2$, then $q_1(x) = q_2(x)$.
\end{theorem}

\begin{proof}
The function $A(\alpha^\prime, \alpha)$ is analytic with respect to
$\alpha^\prime$ and $\alpha$ on the variety (1.8). Therefore its values on
$S^2_1 \times S^2_2$ extend uniquely by analyticity to $M \times M$. In
particular $A(\alpha^\prime, \alpha)$ is uniquely determined in
$S^2 \times S^2$. By (1.30) one gets:
\begin{equation}
        \int_{B_a} p(x) u_1(x, \alpha) u_2 (x, -\alpha^\prime) dx = 0
        \quad \forall \alpha, \alpha^\prime \in S^2.
        \end{equation}

By property $C$ (section 1.2.6), formulas (1.22) -(1.23)) and by 
(1.18), the
orthogonality relation (2.1) implies $p(x) \equiv 0$.
\end{proof}

%%%%%%%%%%%%%%%%%%%%%%%%%%%%%%%%%%%%%%%%%
\subsection{Reconstruction formula for exact data.}  
Fix an arbitrary $\xi \in \R^3$ and choose arbitrary $\theta^\prime, \theta$
satisfying (1.9).

Denote
\begin{equation}
        \widetilde{q} (\xi) := \int_{B_a} e^{-i \xi \cdot x} q(x)dx.
        \end{equation}

Multiply (1.11) by $\nu(\alpha, \theta) \in L^2(S^2)$, where
$\nu(\alpha, \theta)$ will be fixed later, and integrate over $S^2$ with
respect to $\alpha$:
\begin{equation}
        -4\pi \int_{S^2} A(\alpha^\prime, \alpha) \nu(\alpha, \theta)
        d \alpha = \int_{B_a} e^{-i \alpha^\prime \cdot x}
        \int_{S^2} u(x, \alpha) \nu(\alpha, \theta) d \alpha
        q(x)dx.
        \end{equation}

If $q \in Q_a$, then estimates (1.27) and (1.28) imply that the series (1.25)
converges, when $\alpha^\prime$ is replaced by $\theta^\prime \in M$,
uniformly and absolutely on $S^2 \times M_c$. where $M_c \subset M$ is an
arbitrary compact subset of $M$. Formula (1.11) implies that $\alpha^\prime$
can be replaced by $\theta^\prime \in M$, since $B_a$ is a compact set in
$\R^3$.

Define
\begin{equation}
        \rho(x) := \rho(x;\nu) := e^{-i \theta \cdot x} \int_{S^2}
        u(x, \alpha) \nu(\alpha, \theta)d \alpha -1,
        \end{equation}
and rewrite (2.3), with $\alpha^\prime = \theta^\prime$, as
\begin{equation}
 \begin{split}
    -4\pi \int_{S^2} A(\theta^\prime, \alpha) \nu(\alpha, \theta) d \alpha
        &= \int_{B_a} e^{-i \theta^\prime \cdot x + i \theta \cdot x}
          [\rho(x) + 1]q(x)dx \\
        &=\widetilde{q}(\xi)+ \int_{B_a}  e^{-i \xi \cdot x} \rho(x) q(x)dx
          = \widetilde{q} + \varepsilon,
        \end{split}
        \end{equation}
where
\begin{equation}
        |\varepsilon| \leq \Vert q \Vert_a 
        \Vert \rho \Vert_a, \quad
        \Vert q \Vert_a := \Vert q \Vert_{L^2(B_a)}.
        \end{equation}

The following estimate (see \cite{R6}, estimate (2.17) and
its proof in section 6 below) 
holds for a suitable choice of
$\nu(\alpha, \theta)$:
\begin{equation}
        \Vert \rho \Vert_a \leq c |\theta|^{-1}, \quad
        |\theta| \to \infty, \quad \theta \in M.
        \end{equation}

{\it
From (2.5) and (2.7) one gets the reconstruction formula for inversion of
exact fixed-energy 3D scattering data:
\begin{equation}
 \lim_{\substack {
        |\theta| \to \infty \\
        \theta^\prime - \theta = \xi, \\
        \theta,\theta^\prime\in M.} }
 \left\{-4 \pi \int_{S^2} A(\theta^\prime, \alpha) \nu (\alpha, \theta)
    d \alpha \right\} = \widetilde{q}(\xi),
        \end{equation}
and the error estimate:
\begin{equation}
        -4\pi \int_{S^2} A(\theta^\prime, \alpha) \nu(\alpha, \theta)
        d \alpha = \widetilde{q} (\xi) + O\left(\frac{1}{|\theta|}\right),
        \quad |\theta| \to \infty, \quad
        \theta \in M,
        \end{equation}
where (1.9) is always assumed.
}

Let us give an {\it algorithm for computing the function
$\nu(\alpha, \theta)$ for which (2.7), and therefore (2.9), hold, given the
scattering data $A(\alpha^\prime, \alpha)$
$\forall \alpha^\prime, \alpha \in S^2$.}

Fix arbitrarily two numbers $a_1$ and $b$ such that
\begin{equation}
        a < a_1 < b,
        \end{equation}
and define the $L^2$-norm in the annulus:
\begin{equation}
        \Vert \rho \Vert^2 := \int_{a_1 \leq |x| \leq b} |\rho|^2
        dx.
        \end{equation}

Consider the minimization problem
\begin{equation}
        \Vert \rho \Vert = \inf := d(\theta),
        \end{equation}
where the infimum is taken over all $\nu \in L^2(S^2)$.

It is proved in \cite{R6} (see also section (6.3) below) that
\begin{equation}
        d(\theta) \leq c |\theta|^{-1} \hbox{\ if\ } \theta \in M, \quad
        |\theta| \gg 1.
        \end{equation}

The symbol $|\theta| \gg 1$ means that $|\theta|$ is sufficiently large. The
constant $c>0$ in (2.13) depends on the norm $\Vert q \Vert_a$ but
not on the potential $q(x)$ itself. An algorithm for computing a function
$\nu(\alpha, \theta)$, which can be used for inversion of the fixed-energy
3D scattering data by formula (2.9), is as follows:

a) Find any approximate solution to (2.12) in the sense
\begin{equation}
        \Vert \rho (x, \nu) \Vert < 2 d(\theta),
        \end{equation}
where in place of 2 in (2.14) one could put any fixed constant greater
than 1.

b) Any such $\nu (\alpha, \theta)$ generates an estimate of
$\widetilde {q}(\xi)$ with the error $o\left(\frac{1}{|\theta|}\right)$,
$|\theta | \to \infty$.
This estimate is calculated by the formula
\begin{equation}
        \widehat{q} := -4\pi \int_{S^2} A(\theta^\prime, \alpha)
        \nu (\alpha, \theta) d \alpha,
        \end{equation}
where $\nu(\alpha, \theta) \in L^2(S^2)$ is any function satisfying 
(2.14).

We have obtained the following result:

\begin{theorem}
One has
\begin{equation}
        \left|\widehat{q} - \widetilde{q}(\xi) \right| \leq
        \frac{c}{|\theta|}, \quad
        |\theta| \to \infty, \quad
        \theta \in M,
        \end{equation}
provided that (2.14) and (1.9) hold.
\end{theorem}

\begin{proof}
The proof is the same as the proof of (2.5) - (2.7) and is based on the
following estimate \cite{R2}, \cite{R6}:
\begin{equation} 
        \Vert \rho \Vert_a \leq c
        \left(\Vert \rho \Vert + |\theta|^{-1} \right), \quad
        |\theta | \gg  1, \quad \theta \in M.
        \end{equation}
The proof of (2.17) is not simple \cite{R6}. It is given in section 6.
\end{proof}

%%%%%%%%%%%%%%%%%%%%%%%%%%%%%%%%%%%%%%%%%
\subsection{Stability estimate for inversion of the exact data.} 
Let the potentials $q \in Q$, $j= 1,2$, generate the scattering amplitudes
$A_j(\alpha^\prime, \alpha)$.

Let us assume that
\begin{equation}
        \sup_{\alpha^\prime, \alpha \in S^2}
        \left| A_1(\alpha^\prime, \alpha) - A_2(\alpha^\prime, \alpha) \right|
        < \delta.
        \end{equation}

We want to estimate $p(x) := q_1(x) - q_2(x)$.

The main tool is formula (1.30).

The result is (\cite{R6}, \cite{R2}):

\begin{theorem} 
If $q_j \in Q$ and (2.18) holds then
\begin{equation} 
        \sup_{|\xi| < \xi_0} |\widetilde{q_1}(\xi) - \widetilde{q_2}(\xi)|
        \leq c \frac{\ln|\ln \delta|}{|\ln \delta|}, \quad
        \delta \to 0,
        \end{equation}
where $\xi_0 > 0$ is an arbitrary large fixed number and the constant
$c>0$ does not depend on $\delta >0$, $\delta \to 0$.
\end{theorem}

\begin{proof} 
Multiply both sides of (1.30) by
$\nu_1(\alpha, \theta) \nu_2(-\alpha^\prime, \theta_2)$,
where $\theta_j \in M$, $j=1,2,$ $\theta_1 + \theta_2 = \xi$,
$|\theta_1| \to \infty$,
and integrate with respect to $\alpha$ and $\alpha^\prime$ over
$S^2$, to get:
\begin{equation}
 \begin{split}
  &-4\pi \int_{S^2}\int_{S^2} A(\alpha^\prime, \alpha)
   \nu_1(\alpha, \theta_1)
   \nu_2(-\alpha^\prime, \theta_2) d \alpha d \alpha^\prime = \\
  &\int_{B_a} dx p(x) \int_{S^2} u_1(x, \alpha) \nu_1(\alpha, \theta_1)
   d \alpha \int_{S^2} u_2(x, \beta) \nu_2(\beta, \theta_2) d\beta, \quad
   \beta = -\alpha^\prime.
 \end{split}
\end{equation}

Choose $\nu_1$ and $\nu_2$ such that
\begin{equation}
        \Vert \rho_j (\nu_j) \Vert \leq
        \frac{c}{|\theta_j|}, \quad |\theta_j| \to \infty,
        \quad \theta_j \in M,
        \end{equation}
where
$\rho_j(\nu_j) :=\rho_j(x,\nu_j)
:= e^{-i\theta_j x} \int_{S^2} u_j(x, \alpha)
\nu_j(\alpha, \theta_j)d \alpha -1$, and note that
$\frac{|\theta_1|}{|\theta_2|} \to 1$ as $|\theta_1| \to \infty$,
$\theta_1, \theta_2 \in M$, $\theta_1 + \theta_2 = \xi$,
$|\xi| \leq \xi_0$, $\xi_0$ is an arbitrary large but fixed number. From
(2.18), (2.20) and (2.21) one gets
\begin{equation}
        |\widetilde{p}(\xi)| \leq c
        \left(|\theta|^{-1} + c\delta \Vert \nu_1 \Vert_{L^2(S^2)}
        \Vert \nu_2 \Vert_{L^2(S^2)} \right),
        \end{equation}
where $\theta := \theta_1$. One can choose $\nu_1$ and $\nu_2$ such that
(\cite{R6}, see also section 6.7 below)
\begin{equation}
        \Vert \nu_j (\alpha, \theta_j) \Vert_{L^2(S^2)} \leq
        ce^{c|\theta| \ln |\theta|},
        \quad \theta = \theta_1,
        \quad |\theta| \to \infty,
        \quad \theta_2 = \xi -\theta,
        \end{equation}
where $c>0$ stands for various different constants.

Thus (2.22) yields:
\begin{equation}
        \sup_{|\xi| \leq \xi_0} \left|\widetilde{p}(\xi) \right| \leq
        c \min_{s\gg 1} \left[s^{-1} + c_1 \delta e^{c_2s\ln |s|} \right],
        \quad 0 < \delta << 1,
        \end{equation}
where $c, c_1$ and $c_2$ are some positive constants, $s := |\theta|$,
$\delta\ll 1$ means that $\delta>0$ is small and
$s\gg 1$ means that $s>0$ is large. However, our argument is valid for
$s\geq 1$ and $0<\delta\leq \frac{1}{2}$.

One gets
\begin{equation}
        \min_{s>0} \left[s^{-1} + c_1 \delta e^{c_2s\ln|s|}\right] :=
        \eta(\delta) \leq c_3 \frac{\ln |\ln \delta|}{|\ln \delta|},
        \quad \delta \to 0,
        \end{equation}
and the minimizer is
\begin{equation}
        s = s(\delta) =c_2^{-1} \frac{\ln |\delta|}{\ln |\ln \delta|}
        \left[1 + o(1) \right], \quad \delta \to 0.
        \end{equation}
From (2.24)-(2.26) one gets (2.19).
\end{proof}

\begin{remark}
In the above proof the difficult part is the proof of (2.23). Estimate
(2.23)
can be derived for $\nu(\alpha, \theta)$ such that
\begin{equation}
        \Vert \psi (x, \theta) - \int_{S^2} u(x, \alpha)
        \nu(\alpha, \theta) d \alpha \Vert_{L^2(B_b)}
        \leq
        c \frac{e^{-b \kappa}}{\kappa}, \quad
        \kappa := |Im \theta|, \ \theta \in M, \ \theta \gg 1,
        \end{equation}
and $\Vert \nu(\alpha, \theta) \Vert_{L^2(S^2)} =\inf$,
where the infimum is taken over all $\nu \in L^2(S^2)$.
\end{remark}

In section 6.7 we
consider the problem of finding $\nu \in L^2(S^2)$ with minimal norm
$\Vert \nu \Vert_{L^2(S^2)} := a(\nu)$
among all $\nu(\alpha, \theta)$ which satisfy the inequality:
\begin{equation}
        \Vert \psi(x) - \int_{S^2} u(x, \alpha) \nu (\alpha, \theta)
        d \alpha \Vert_{L^2(B_b)} \leq \varepsilon.
        \end{equation}

The necessity to consider the $\nu$ with the minimal norm
$\Vert \nu \Vert_{L^2(S^2)}$ comes from the simple observation:
there exists a sequence of $\nu_n \in L^2(S^2)$,
$\Vert \nu_n \Vert_{L^2(S^2)} = 1$, such that
\begin{equation}
        \Vert \int_{S^2} u(x, \alpha) \nu_n(\alpha) d\alpha
        \Vert_{L^2(B_b)} \to 0 \hbox{\ as\ } n \to \infty.
        \end{equation}
To prove (2.29) note that
\begin{equation}
        u(x, \alpha) = e^{i \alpha \cdot x} -\int_{B_b}
        \frac{e^{i|x-y|}}{4 \pi|x-y|} q(y) u(y, \alpha) dy :=
        e^{i \alpha \cdot x} - Tu,
        \end{equation}
where the operator $(I +T)^{-1} := I+T_1$ is a continuous bijection of
$C(B_b)$ onto itself, and $C(B_b)$ is the usual space of continuous in
$B_b$, $b \geq a$, functions
equipped with the sup-norm \cite{R6}. Since $T$ is compact in
$C(B_b)$, the above statement follows from the injectivity of $I+T$,
which we now prove: 

If $f +Tf =0$, then $f$ is extended to $C(\R^3)$ by
the formula $f = -Tf$, and satisfies the following equation
$(\nabla^2 + 1 - q(x))f = 0$ in $\R^3$ and the radiation condition of the
type (1.6) with $k=1$.
Therefore $f(x) \equiv 0$ and the injectivity of $I+T$ is proved.

 Thus
$I+T$ and $I +T_1$ are continuous bijections of $C(B_b)$ into itself for
any $b \geq a$.

Writing $u(x, \alpha) = (I +T_1)e^{i \alpha \cdot x}$, one concludes that
(2.29) is equivalent to
\begin{equation}
        \Vert \int_{S^2} e^{i \alpha \cdot x} \nu_n(\alpha) d \alpha
        \Vert_{L^2(B_b)} \to 0 \hbox{\ as\ }
        n \to \infty.
        \end{equation}
Existence of a normalized sequence $\nu_n(\alpha)$ satifying (2.31)
follows from the compactness of the operator
$$Q : L^2(S^2) \to L^2(B_b), \quad Q \nu := \int_{S^2} e^{i \alpha \cdot x}
\nu(\alpha) d \alpha.$$
Of course, the same argument is applicable to the operator
$Q_1\nu := \int_{S^2} u(x, \alpha) \nu(\alpha) d \alpha$, but the
bijectivity of $I+T$ in $C(B_b)$, $b \geq a$, is of independent interest.

It follows from (2.29) that, for a given $\varepsilon >0$, one can find
$\nu$ in (2.29) with an arbitrary large norm
$\Vert \nu \Vert_{L^2(S^2)}$. By this reason we are interested in
$\nu$ with minimal norm. Estimate (2.23) gives a bound on the growth of the
minimal value of the norm $\Vert \nu \Vert_{L^2(S^2)}$, where
$\nu = \nu(\alpha, \theta)$ satisfies (2.28) with
$\varepsilon = \frac{e^{-b \kappa}}{\kappa}$, $ \kappa := |Im \theta|$,
$|\theta| \to \infty$, $\theta \in M$.

%%%%%%%%%%%%%%%%%%%%%%%%%%%%%%%%%%%%%%%%%
\subsection{Stability estimate for inversion of noisy data.}  
Assume now that the scattering data are given with some error: a function
$A_\delta (\alpha^\prime, \alpha)$ is given such that
\begin{equation}
        \sup_{\alpha^\prime, \alpha \in S^2}
        \left|A(\alpha^\prime, \alpha) -
        A_\delta(\alpha^\prime, \alpha) \right| \leq \delta.
        \end{equation}

We emphasize that $A_\delta (\alpha^\prime, \alpha)$ is not necessarily a
scattering amplitude corresponding to some potential, it is an arbitrary
function in $L^\infty (S^2 \times S^2)$ satisfying (2.32). It is assumed that
the unknown function $A(\alpha^\prime, \alpha)$ is the scattering amplitude
corresponding to a $q \in Q$.

The problem is: {\it Give an algorithm for calculating $\widehat{q}$
such that}
\begin{equation}
        \sup_{|\xi| \leq \xi_0} \left|\widehat{q} - \widetilde{q} (\xi)
        \right| \leq \eta(\delta), \quad \eta(\delta) \to 0
        \hbox{\ as\ } \delta \to 0,
        \end{equation}
{\it where $\xi_0 >0$ is an arbitrary large fixed number, and estimate
the rate at which $\eta(\delta)$ tends to zero.}

An algorithm for inversion of noisy data will now be described.

Let
\begin{equation} 
        N(\delta) := \left[ \frac{|\ln \delta|}{\ln|\ln \delta|} \right],
        \end{equation}
where $[x]$ is the integer nearest to $x>0$,
\begin{equation}
        \widehat{A}_\delta (\theta^\prime, \alpha) :=
        \sum^{N(\delta)}_{\l =0} A_{\delta \l} (\alpha) Y_\l
        (\theta^\prime), \quad
        A_{\delta \l} (\alpha) := \int_{S^2}
        A_\delta (\alpha^\prime, \alpha) \overline{Y_\l(\alpha^\prime)}
        d \alpha^\prime,
        \end{equation}
\begin{equation}
        u_\delta (x, \alpha) := e^{i \alpha \cdot x} +
        \sum^{N(\delta)}_{\l =0} A_{\delta \l}
        (\alpha) Y_\l (\alpha^\prime) h_\l (r),
        \end{equation}
\begin{equation} 
        \rho_\delta (x; \nu) := e^{-i \theta \cdot x} \int_{S^2}
        u_\delta (x, \alpha) \nu (\alpha) d \alpha -1, \quad
        \theta \in M,
        \end{equation}
\begin{equation}
        \mu(\delta) := e^{-\gamma N(\delta)}, \quad
        \gamma = \const > 0,
        \end{equation}
\begin{equation} 
        a(\nu) := \Vert \nu \Vert_{L^2(S^2)}, \quad
        \kappa := |Im \theta|.
        \end{equation}

Consider the variational problem with constraints:
\begin{equation}
        |\theta| = \sup := \vartheta(\delta),
        \end{equation}
\begin{equation}
        |\theta| \left[ \Vert \rho_\delta (\nu) \Vert +
        a(\nu) e^{\kappa b} \mu(\delta) \right] \leq c,
        \quad \theta \in M,
        \end{equation}
where the norm is defined in (2.11), and it is assumed that
\begin{equation}
        \theta^\prime - \theta = \xi, \quad
        \theta,\theta^\prime \in M,
        \end{equation}
where $\xi \in \R^3$ is an arbitrary fixed vector, $c>0$ is a sufficiently
large constant, and the supremum is taken over $\theta \in M$ and
$\nu \in L^2(S^2)$ under the constraints (2.41).

Given $\xi \in \R^3$ one can always find $\theta$ and $\theta^\prime$
such that (2.42) holds.

We prove that $\vartheta(\delta) \to \infty$, in fact
\begin{equation}
        \vartheta(\delta) \geq c \frac{|\ln \delta|}{(\ln|\ln \delta|)^2},
        \quad \delta \to 0.
        \end{equation}
Let $\theta(\delta)$, $\nu_\delta (\alpha)$ be any approximate solution to
problem (2.40)-(2.41) in the sense that
\begin{equation}
        |\theta(\delta)| \geq \frac{\vartheta(\delta)}{2}.
        \end{equation}
Calculate
\begin{equation}
        \widehat{q}_\delta := -4 \pi \int_{S^2} \widehat{A}_\delta
        (\theta^\prime, \alpha) \nu_\delta (\alpha) d \alpha.
        \end{equation}

\begin{theorem} 
One has
\begin{equation}
        \sup_{|\xi| \leq \xi_0} \left| \widehat{q}_\delta -
        \widetilde{q}(\xi) \right| \leq c
        \frac{(\ln |\ln \delta|)^2}{|\ln \delta|} \hbox{\ as\ }
        \delta \to 0.
        \end{equation}
\end{theorem}

\begin{proof}
\begin{equation}
  \begin{split}
     \widehat{q_\delta} - \widetilde{q} (\xi)
        = & -4\pi \int_{S^2} \widehat{A}_\delta (\theta^\prime, \alpha)
            \nu_\delta (\alpha) d \alpha - \widetilde{q} (\xi)   \\
        = & -4\pi \int_{S^2} A(\theta^\prime, \alpha)
            \nu_\delta(\alpha) d \alpha - \widetilde{q}(\xi)      \\
          & + 4\pi \int_{S^2} [A(\theta^\prime, \alpha) -
           \widehat{A}_\delta (\theta^\prime, \alpha)]
           \nu_\delta (\alpha) d \alpha.
   \end{split}
   \end{equation}

The rest of the proof consists of the following steps:

\underline{Step 1.} We prove that
\begin{equation} 
        \Vert \rho(x, \nu_\delta) \Vert \leq
       c[ \Vert \rho_\delta (x, \nu_\delta)\Vert +
        a(\nu_\delta) e^{\kappa b} \mu(\delta)] \leq
        \frac{c}{\vartheta(\delta)},
        \end{equation}
where the norm is defined in (2.11).

This estimate and (2.43) imply (see the proof of (2.19) and (2.26)) that
\begin{equation}
        \left| -4\pi \int_{S^2} A(\theta^\prime, \alpha) \nu_\delta (\alpha)
        d \alpha - \widetilde{q}(\xi) \right| \leq
        c \frac{(\ln|\ln \delta|)^2}{|\ln \delta|}.
        \end{equation}

\underline{Step 2.} We prove that
\begin{equation}
        \left| \int_{S^2} \left[A(\theta^\prime, \alpha) - \widehat{A}_\delta
        (\theta^\prime, \alpha \right] \nu_\delta (\alpha) d \alpha \right|
        \leq ca(\nu_\delta) e^{\kappa b} \mu(\delta) \leq
        \frac{c}{\vartheta(\delta)} \leq c
        \frac{(\ln|\ln \delta|)^2}{|\ln \delta|},
        \end{equation}
where $\theta^\prime = \theta^\prime(\delta) = \xi + \theta(\delta)$,
and the pair $\{\theta(\delta), \nu_\delta (\alpha)\}$ solves
(2.40)-(2.41) approximately in the sense specified above.
(See formula (2.44)).

This estimate follows from (2.41) and from the inequality
\begin{equation}
        \Vert A(\theta^\prime, \alpha) - \widehat{A}_\delta
        (\theta^\prime, \alpha) \Vert_{L^2(S^2)} \leq c e^{\kappa b}
        \mu(\delta).
        \end{equation}

Let us prove (2.51). One has
\begin{equation}
   \begin{split}
     & \Vert \widehat{A}_\delta (\theta^\prime(\delta), \alpha)
            - A(\theta^\prime (\delta), \alpha)
       \Vert_{L^2(S^2)} \leq
       \Vert \sum^{N(\delta)}_{\l =0}
          \left[\widehat{A}_{\delta \l} (\alpha) - A_l(\alpha) \right]
          Y_\l (\theta^\prime)
       \Vert_{L^2(S^2)} \\
    &\qquad +
       \Vert \sum^\infty_{\l = N(\delta) +1}
          A_\l (\alpha)
          Y_\l(\theta^\prime)
        \Vert_{L^2(S^2)} := I_1 + I_2,
        \end{split}
        \end{equation}
where $N(\delta)$ is given in (2.34) and
$\theta^\prime := \theta^\prime(\delta)$.
Using (1.28), (1.29) one gets
\begin{equation}
        I_1 \leq c\delta N^2(\delta)
        \frac{e^{\kappa b}(2N)^{N(\delta)+1}}{(eb)^{N(\delta)}}.
        \end{equation}
Here we have used the estimate
$$\sup_{\alpha \in S^2} \sum^\infty_{\l =0} \left|A_\l(\alpha) -
  \widehat{A}_{\delta \l}(\alpha) \right|^2 \leq 4 \pi \delta^2,$$
which follows from (2.32) and the Parseval equality, and implies
\begin{equation}
        \sup_{\alpha \in S^2} \left|A_\delta (\alpha) -
        \widehat{A}_{\delta \l} (\alpha) \right| \leq \sqrt{4 \pi} \delta.
        \end{equation}

We also took into account that there are $(N + 1)^2$ spherical harmonics
$Y_\l = Y_{\l,m}$  with $0 \leq \l \leq N$, because
$\sum^\l_{m = -\l} 1 = 2 \l +1$, and $\sum^N_{\l=0} (2\l +1) = (N +1)^2$.
For large $N$ one has $(N+1)^2 = N^2[1+o(1)]$, $N \to \infty$, so we
write $(N+1)^2 \leq cN^2$, $c>1$.

To estimate $I_2$, use (1.27) -(1.29) and get:
\begin{equation}
        I_2 \leq c \sum^\infty_{\l = N+1}
        \left(\frac{ea}{2\l}\right)^{\l+1} 
        \frac{e^{\kappa a_1}}
        {\left(\frac{e a_1}{2\l}\right)^{\l+1}} \leq
        ce^{\kappa a_1}
        \left(\frac{a}{a_1}\right)^{N+1}_.
        \end{equation}

Minimizing with respect to $N>1$ the function
\begin{equation}
        \delta N^2 \frac{(2N)^{N+1}}{(eb)^N} + s^{N+1}, \quad
        0 <s := \frac{a}{a_1} < 1.
        \end{equation}
one gets
\begin{equation}
        \min_{N>1} \left[\delta N^2 \frac{(2N)^{N+1}}{(eb)^N} + s^{N+1}
        \right]
        \leq ce^{-\gamma N(\delta)} = c \mu (\delta), \quad
        \gamma = \ln \frac{a_1}{a} > 0,
        \end{equation}
where $N(\delta)$ is given in (2.34) and $\mu(\delta)$ is defined by
(2.38). Thus, from (2.52) -(2.56) one gets (2.51). Theorem 2.4 is proved.
\end{proof}

%%%%%%%%%%%%%%%%%%%%%%%%%%%%%%%%%%%%%%%%%
\subsection{Stability estimate for the scattering solutions.}  
Let us assume (2.18) and derive the following estimate:

\begin{theorem} 
If $q_1, q_2 \in Q_a$ and (2.18) holds then
\begin{equation}
        \sup_{\substack{x \in B^\prime_a \\ \alpha \in S^2}}
        |u_1(x, \alpha) - u_2(x, \alpha)| \leq c \mu(\delta), \quad
        |x| >a,
        \end{equation}
where $\mu(\delta)$ is defined by (2.38) and (2.34), and $c>0$ is a
constant.
\end{theorem}

\begin{proof}
Using (1.31), one gets:
\begin{equation}
        u_1 - u_2 = \sum^\infty_{\l =0} \left[A_{\l 1} (\alpha) -
        A_{\l 2} (\alpha) \right] Y_\l (\alpha^\prime) h_\l (r),
        \quad r >a.
        \end{equation}

As stated below formula (1.33), one has
\begin{equation}
        \left| h_\l (r) \right| \leq
        \left| h_{\l +1} (r) \right|, \quad
        r >0.
        \end{equation}

From (2.58) one gets:
\begin{equation}
        \Vert u_1 - u_2
        \Vert^2_{L^2(B_b \backslash B_{a_1})} =
        \sum^\infty_{\l =0} |A_{\l 1} (\alpha) - A_{\l 2} (\alpha) |^2
        \int^b_{a_1} r^2|h_\l(r)| dr, \quad
        a < a_1 < b.
        \end{equation}

It follows from (2.60) and (1.34) that
\begin{equation}
    \begin{split}
        \sup_{0 \leq \l \leq N} \int^b_{a_1} r^2 |h_\l(r)|^2 dr =
        \int^b_{a_1} r^2|h_N(r)|^2 dr \leq
    \\
        c \left(\frac{2N}{e}\right)^{2N} \int^b_{a_1}
        \frac{dr}{r^{2N}} \leq c \left(\frac{2N}{e} \right)^{2N}
a_1^{-2N},\quad
        b>a_1.
    \end{split}
    \end{equation}
From (2.60)-(2.61) one gets
\begin{equation}
       \Vert u_1 - u_2 \Vert^2_{L^2(B_b \backslash B_a)}
       \leq
        c \delta^2 \left(\frac{2N}{ea_1} \right)^{2N}
         + c \sum^\infty_{\l = N+1}
        \left(|A_{\l 1}|^2 + |A_{\l 2}|^2 \right)
        \left(\frac{2\l}{ea_1} \right)^{2\l} ,
        \end{equation}
where we have used the monotone decrease of $r|h_\l(r)|^2$ as a function of
$r$. Using estimate (1.27)in order to estimate
$|A_{\l j} (\alpha)|$, $j = 1,2,$ one gets:
\begin{equation}
        \Vert u_1 - u_2 \Vert^2_{L^2(B_b \backslash B_{a_1})} \leq
        c \delta^2 \left(\frac{2N}{ea_1}\right)^{2N}
         + c \left(\frac{a}{a_1}\right)^{2N}, \quad
        a < a_1.
        \end{equation}
Minimization of the right-hand side of (2.63) with respect to $N \geq 1$
yields, as in (2.56), the estimate similar to (2.56):
\begin{equation}
        \Vert u_1 - u_2 \Vert^2_{L^2(B_b \backslash B_{a_1})} \leq
        ce^{-\gamma N(\delta)}, \quad
        \gamma = \ln \frac{a_1}{a} > 0.
        \end{equation}

Since $u_1 - u_2 := w$ solves the equation
\begin{equation}
        (\nabla^2 +1)w=0 \hbox{\ in\ } B_a^\prime := \R^3 \backslash B_a,
        \end{equation}
one can use the known elliptic estimate:
\begin{equation}
        \Vert w \Vert_{H^2(D_1)} \leq c \left(\Vert (\nabla^2 +1)
        w \Vert_{L^2(D_2)} + \Vert w \Vert_{L^2(D_2)} \right),
        \quad
        D_1 \subset D_2,
        \end{equation}
where $D_1$ is a strictly inner subset of $D_2$ and $c=c(D_1, D_2)$,
and get:
\begin{equation}
        \Vert w \Vert_{H^2(D_1)} \leq ce^{-\gamma N(\delta)},
        \end{equation}
where $D_1$ is any annulus $a_1 < a_2 \leq |x| \leq a_3 < b$. By the embedding
theorem, (2.68) implies (2.58) in $\R^3$.
\end{proof}

%%%%%%%%%%%%%%%%%%%%%%%%%%%%%%%%%%%%%%
\subsection{Spherically symmetric potentials.}  
If $q(x) = q(r)$, $r := |x|$, then
\begin{equation}
        A(\alpha^\prime, \alpha) = A(\alpha^\prime \cdot \alpha), \quad
        A_\l(\alpha)= A_\l Y_\l (\alpha).
        \end{equation}
In \cite{R16}
{\it
the converse is proved: if $q \in Q_a$ and (2.69)
holds then $q(x) = q(r)$.
}

The scattering data $A(\alpha^\prime, \alpha)$ in the case of the spherically
summetric potential is equivalent to the set of the phase shifts $\delta_\l$.
The phase shifts are defined as follows:
\begin{equation}
        1 + \frac{i}{2 \pi} A_\l = e^{2i \delta_\l}, \quad
        A_\l = 4 \pi e^{i \delta_\l} \sin \delta_\l, \quad k = 1.
        \end{equation}
From Theorem 2.1 it follows that is $q=q(r) \in Q_a$ then the set
$\{\delta_\l\}_{\l=0,1,2,\dots}$ determines uniquley $q(r)$. A much stronger
result is proved by the author in \cite{R17}. To formulate this result,
denote by ${\mathcal L}$ any subset of nonnegative integers such that
\begin{equation}
        \sum_{\l \in {\mathcal L}, \l \neq 0} \frac{1}{\l} = \infty.
        \end{equation}

\begin{theorem}(\cite{R17})
If $q(x) = q(r) \in Q_a$ then the data $\{\delta_\l\}_{\forall \l \in L}$
determine $q(r)$ uniquely.
\end{theorem}

In \cite{R10} an example is given of two quite different potentials
$q_1(r)$ and $q_2(r)$, piecewise-constant, $q_j(r) = 0$ for
$r>5$, for which
$$\sup_{\l \geq 0} \left|\delta^{(1)}_\l - \delta^{(2)}_\l \right|<10^{-5}
  \hbox{\ and\ } \sup_{r \geq 0} \left|q_1(r) -q_2(r) \right| \geq 1,
  \hbox{\ where\ }q_1 \hbox{\ and\ } q_2$$
are of order of magnitude of 1.

This result shows that the stability estimate (2.19) is accurate.

%%%%%%%%%%%%%%%%%%%%%%%%%%%%%%%%%%%%%%%%%
\section{The direct obstacle scattering problem with fixed-frequency
data}%3
Let $D \subset \R^3$ be a bounded domain (an obstacle) with  boundary
$S$. We want to study the obstacle scattering problem under minimal
smoothness assumptions on $S$.

Recall that if $S$ is $C^{1, \lambda}$, $0 < \lambda \leq 1$, that is,
in local coordinates $S$ is a graph of a $C^{1, \lambda}$- function
$x_3 = g(x^\prime)$, $x^\prime := (x_1, x_2)$, $g \in C^{1, \lambda}$,
then the obstacle scattering problem consists of finding the scattering
solution $u(x, \alpha)$, which satisfies the equations:
\begin{equation}
        (\nabla^2 +1) u = 0 \hbox{\ in\ } D^\prime := \R^3 \backslash D,
        \end{equation}
\begin{equation}
        \Gamma u = 0,
        \end{equation}
\begin{equation}
        u = e^{i \alpha \cdot x} + A(\alpha^\prime, \alpha)
        \frac{e^{ir}}{r} + o\left(\frac{1}{r}\right), \quad
        r := |x| \to \infty, \quad \frac{x}{r} = \alpha^\prime,
        \end{equation}
where $\alpha \in S^2$ is given and (3.2) is the Dirichlet condition if
$\Gamma u =u$, the Neumann condition if $\Gamma u = u_N$, or the Robin
condition if $\Gamma u := u_N + \sigma(s) u$, $\sigma (s) \in L^\infty(S)$,
$Im \sigma (s) = 0$, $N$ is the exterior normal to $S$. We took the
wavenumber $k=1$ without loss of generality.

If $S$ is very rough (non-smooth),
$N$ may be not defined on $S$. The minimal assumptions
on the smoothness of $S$, under which the existence and uniqueness of the
solution to the direct scattering problem are established, were introduced in
\cite{R8} and \cite{R9}. These assumptions are:

$A_1$) If $\Gamma u =u$ then $D$ is an
arbitrary bounded domain, that is a bounded open set.
       
$A_2$) If $\Gamma u = u_N$ then the assumption on $S$ is:
\begin{equation}
        i : H^1(D^\prime_a) \to L^2(D^\prime_a)
        \hbox{\ is\ } \hbox{\ compact\ }, \quad
        D^\prime_a := D^\prime \cap B_a.
        \end{equation}
Here $i$ is the embedding operator, $H^1$ is the Sobolev space,
$a>0$ is such a number that the ball $B_a := \{x: |x| \leq a\}$ contains
$D$, and $D^\prime := \R^3 \backslash D$.

$A_3$) If $\Gamma u=u_N + \sigma(s) u$, then the assumption on $S$ is:
\begin{equation}
        i : H^1(D_{a^\prime}) \to L^2(D_{a^\prime}) \hbox{\ and\ }
        i_1 : H^1(D_{a^\prime}) \to L^2(S) \hbox{\ are\ } \hbox{\ compact.}
        \end{equation}

Here the integration on $S$ in the definition of $L^2(S)$ is understood with
respect to the two-dimensional Hausdorff measure on $S$.

The usual classes of domains in the theory of Sobolev spaces are:

1) domains satisfying the cone condition,

2) Lipschitz domains,

\nd and

3) extension domains, (see \cite{M}, \cite{MP}).

They all are such that the above assumptions $A_2$) and $A_3$) hold. Let us recall
the definitions of these domains: $D$ satisfies the cone condition if each
point of $D$ is the vertex of a cone contained in $D$ along with its
closure, the cone in the local coordinates is the region
$x^{\prime 2} < ax^2_3$, $0 < x_3 <b$, and $a,b>0$ are fixed positive
constants. A domain
$D$ is Lipschitz if each point of $S:= \partial D$ has a neighborhood
$U \subset \R^n$, such that $U \cap D$ can be mapped onto a cube by
a quasi-isometric map.

A homeomorphic $f:D_1 \to D_2$ is called quasi-isometric if
$$\lim_{x \to x_0} \sup \frac{|f(x) - f(x_0)|}{|x-y_0|} \leq M, \quad
   \lim_{y \to y_0} \sup \frac{|f^{-1}(y) - f^{-1}(y_0)|}{|y-y_0|} \leq M
   $$
for any $x_0 \in D_1$ and any $y_0 \in D_2$ and the Jacobian
 $\det f'(x)$ preserves its sign in $D_1$. 

A domain $D$ is an extension domain in $H^1$ if there exists a linear
continuous operator $E: H^1(D) \to H^1(\R^3)$, $Eu = u$ on $D$ for all
$u \in H^1(D)$.

An extension domain may fail to satisfy the cone condition, but a bounded
domain satisfying the cone condition is the union of a finite number of
extension domains.

We will use also the domains with finite perimeter. A domain $D$ has
finite perimeter
if and only if $\Vert \nabla \chi_D(x) \Vert_{B V(\R^3)} < \infty$,
where $\chi_D(x)$ is the characteristic function of $D$ and $BV$ is the
space of functions of bounded variation \cite{M}.

Let us state the result from \cite{R9}.

\begin{theorem} 
Problem (3.1) - (3.3) has a unique weak solution if:

a) $D$ is an arbitrary bounded domain (open set) in $\R^3$ and
$\Gamma u=u$,

b) $D$ is a bounded domain, condition $A_2$) holds and $\Gamma u=u_N$

c) $D$ is a bounded domain, condition
$A_3$) holds and $\Gamma u=u_N + \sigma (s)u$, $\sigma (s) \in L^\infty (S)$,
$Im \sigma (s) = 0$.
\end{theorem}

The solution to (3.1) - (3.3)
for rough boundaries is understood in the weak sense and
has the following properties:

a) if
$\Gamma u= u$ then $u \in \oH1 (D^\prime_R)\cap C^\infty_{\l oc}(D^\prime)$,
$u\in L^2 \left(D^\prime, \frac{1}{1+|x|^a}\right)$, $a>1$,
and (3.3) holds with
\begin{equation}
        A(\alpha^\prime, \alpha) = -\frac{1}{4\pi} \int_S
        e^{-i \alpha^\prime \cdot s}
        u_N(s, \alpha) ds,
        \end{equation}
where $\oH1 (D^\prime_R)$ is 
the Sobolev space of functions which vanish on $S$,

b) if $\Gamma u = u_N$, then $u \in H^1(D^\prime_R) \cap C^\infty_{\l oc}
(D^\prime)$,
$u\in L^2 \left(D^\prime, \frac{1}{1+|x|^a}\right)$, $a>1$,
and (3.3) holds with
\begin{equation}
        A(\alpha^\prime, \alpha) = \frac{1}{4 \pi} \int_S
        \frac{\partial e^{-i \alpha^\prime \cdot s}}{\partial N}
        u(s, \alpha) ds,
        \end{equation}

c) if $\Gamma u=u_N + \sigma(s) u$, then $u \in H^1(D^\prime_R) \cap
C^\infty_{\l oc} (D^\prime)$,
$u\in L^2 \left(D^\prime, \frac{1}{1+|x|^a}\right)$, $a>1$,
and (3.3) holds with
\begin{equation}
         A(\alpha^\prime, \alpha) = \frac{1}{4 \pi} \int_S
        \left[\frac{\partial e^{-i \alpha \cdot s}}{\partial N_s} +
        \sigma(s)e^{-i \alpha^\prime \cdot s} \right] u(s, \alpha) ds.
        \end{equation}

In \cite{R8} and \cite{R9} Theorem 3.1 was proved with the operator
$\nabla^2$ replaced by a general second-order selfadjoint elliptic
operator. The weak solution is defined in the case
$\Gamma u = u_N + \sigma(s) u$, as a function
$u \in H^1(D^\prime_R) \cap C^\infty_{\l oc} (D^\prime)$ for any
$R > R_0$, $B_{R_0} \supset D$, which satisfies (3.3) and satifies the
integral relation:
\begin{equation}
        \int_{D^\prime} (\nabla u \nabla \phi - u \phi) dx -
        \int_S \sigma(s) u \phi ds = 0 \quad
        \forall \phi \in H^1_0(D^\prime).
        \end{equation}
Here $H_0^1(D^\prime)$ is the set of $H^1(D^\prime)$ functions vanishing
near infinity and $ds$ is the two-dimensional
Hausdorff measure on $S$. Formula (3.9)
makes sense for domains $D$ with finite perimeter.

%%%%%%%%%%%%%%%%%%%%%%%%%%%%%%%%%%%%%%
\subsection{Uniqueness theorem for inverse obstacle scattering.}  
The inverse obstacle scattering problem consists of finding $S$ and the
boundary condition on $S$ given
$A(\alpha^\prime, \alpha) \ \forall \alpha^\prime \in S^2$.
The scattering amplitude $A(\alpha^\prime, \alpha)$ in the obstacle scattering
problem satisfies conditions listed in section 1.2.2.
In particular, it admits
unique analytic continuation from $S^2 \times S^2$ onto $M \times M$.

Let us outline the proof of the uniqueness theorem for inverse obstacle
scattering problem with fixed-frequency data. This theorem belongs to the
author \cite{R7}, but we give a new proof \cite{R12}, see also \cite{R11}.

\begin{theorem}
If $A(\alpha^\prime, \alpha) = A_2(\alpha^\prime, \alpha)$
$\forall \alpha^\prime, \alpha,$ running through arbitrary small open subsets
of $S^2$, then $D_1 = D_2 := D$, and the boundary condition on
$S := \partial D$ are uniquely determined.
\end{theorem}

\begin{proof}
As in the proof of Theorem 2.1, the data determine uniquely
$A_j(\alpha^\prime, \alpha)$ on $S^2 \times S^2$ so that
$A_1(\alpha^\prime, \alpha) = A_2(\alpha^\prime, \alpha) \ \forall
\alpha^\prime, \alpha \in S^2$.
If one has already proved that $D_1 = D_2$,
that is, $S_1 = S_2 := S$, then the boundary condition on $S$ is
uniquely determined because the scattering solution $u(x, \alpha)$ is
uniquely determined by the scattering amplitude in $D^\prime$
(and is analytically determined by formula (1.31) in $B_{a^\prime}$).
Thus, the limiting values of $\frac{u_N}{u}$ on $S$ are uniquely
determined.
If the limit
is zero (almost everywhere on $S$) then $\Gamma u=u_N$, if the limit is
infinity (almost everywhere on $S$) then $\Gamma u=u$, and if the limit is a
function $-\sigma(s)$, then $\Gamma u =u_N + \sigma (s)u$.
Therefore the main
point is to prove that $S$ is uniquely determined by
$A(\alpha^\prime, \alpha) \ \forall \alpha^\prime, \alpha \in S^2$.
Let us prove this.

Assume the contrary: $S_1 \neq S_2$. Let $D_{12} := D_1 \cup D_2$,
$S_{12} = \partial D_{12}$, $D^{12} := D_1 \cap D_2$.
Denote by $\widetilde{D}_1$ a connected component of $D_{12} \backslash D_2$.
We want to show that $D_{12} \backslash D_2$ is an empty set.
An important tool is the formula \cite{R12} similar to (1.30):
\begin{equation}
        \begin{split}
        &-4\pi \left[A_1(\alpha^\prime, \alpha) - A_2(\alpha^\prime, \alpha)
        \right] = \\
        &\int_{S_{12}} \left[ u_{1N}(s, -\alpha^\prime) u_{2}(s, \alpha)-
        u_1(s, -\alpha^\prime) u_{2N}(s, \alpha) \right] ds.
        \end{split}
        \end{equation}
This formula holds for domains with finite perimeter.

If $A_1 = A_2 \ \forall \alpha^\prime, \alpha \in S^2$, then (3.10)
yields:
\begin{equation}
        \int_{S_{12}}\left[u_{1N}(s,- \alpha^\prime) u_2(s, \alpha) -
        u_1(s, -\alpha^\prime) u_{2N}(s, \alpha) \right] ds =0
        \quad
        \forall \alpha^\prime, \alpha \in S^2.
        \end{equation}

From (3.11) and (1.7) one derives:
\begin{equation}
        \int_{S_{12}} \left[G_{1N}(x, s) G_2(s,y) - G_1(x, s)
        G_{2N}(s,y)\right] ds =0 \quad
        \forall x,y \in D^\prime_{12}.
        \end{equation}

From (3.12) and Green's formula one gets:
\begin{equation}
        G_1(x,y) = G_2(x,y) \quad
        \forall x, y \in D^\prime_{12}.
        \end{equation}

This leads to a contradiction unless $D_1 = D_2$.

Indeed,
if $D_1\not= D_2$, i.e., 
if $\widetilde{D_1}$ is not empty, take a point $s$ on the boundary
of $\widetilde{D_1}$ which belongs to $S_1$. This point is an interior
point for $D^\prime_2$. Thus
\begin{equation}
        G_2 (s,y) \to \infty \hbox{\ as\ } y \to s.
        \end{equation}

On the other hand, since $s \in S_1$ one has
$\Gamma G_1(s,y)=0$, i.e.,
$G_1(s,y)=0$ if $\Gamma u=u$,
$G_{1N}(s,y) = 0$ if $\Gamma u =u_N$, $G_{1N}(s,y) + \sigma(s) G_1(s,y) =0$
if $\Gamma u = u_N + \sigma (s) u$. In all the three cases
\begin{equation}
        \Gamma G_2(s,y) \to \infty, \hbox{\ as\ } y \to s,
        \end{equation}
while
\begin{equation}
        \Gamma G_1(s,y) = 0.
        \end{equation}

From (3.13), (3.15) and (3.16) one gets a contradiction.
Theorem 3.1 is proved.
\end{proof}

\begin{remark}
In \cite{R18} a uniqueness theorem is proved for inverse obstacle problem
with the transmission boundary condition: the boundary condition and the
boundary and the wavenumber in the interior of the obstacle are uniquely
determined by the fixed-frequency scattering data $A(\alpha^\prime, \alpha)$
$\forall \alpha^\prime, \alpha \in S^2$.
\end{remark}

\begin{remark} 
The basic reason to consider the domains with finite perimeter is the
validity of the Green formula in such domains \cite{M}.

In \cite{R11} it is proved that is $D_1$ and $D_2$ are domains with finite
perimeter then $\widetilde{D_1}$ is also such a domain.
\end{remark}

%%%%%%%%%%%%%%%%%%%%%%%%%%%%%%%%%%%%%%
\subsection{Stability estimate for inverse obstacle scattering.}
Consider the scattering problem (3.1) - (3.3) with
$\Gamma u =u$, for example. Let $D_1$ and $D_2$ be two arbitrary
obstacles in the class ${\mathcal O}^\lambda$
which consists of the bounded domains
whose boundaries can be covered by finitely many balls $B_j$, on the patch
$S_j = S \cap B_j$ the boundary is described in the local coordinates by
the equation $x_3 = g_j(x^\prime)$, $x^\prime = (x_1, x_2)$,
where $g_j(x^\prime) \in C^{2, \lambda}$, $0 < \lambda \leq 1$, and
\begin{equation}
        \sup_j \Vert g_j \Vert_{C^{2, \lambda}(S_j)} \leq c_0,
        \end{equation}
where $c_0$ does not depend on the choice of $D\in{\mathcal O}^\lambda$.

Let $A_j(\alpha^\prime, \alpha)$ be the scattering amplitude corresponding to
the obstacle $D_j$, $j=1,2$.

Assume that (2.18) holds. Define the symmetric Hausdorff distance between
$S_1 = \partial D_1$ and $S_2 = \partial D_2$:
\begin{equation}
        \rho := \rho(\delta) :=
        \max \{ \sup_{x \in S_2} \inf_{y \in S_1} |x-y|, \quad
        \sup_{x \in S_1} \inf_{y \in S_2} |x-y| \}.
        \end{equation}
The basic stability estimate \cite{R19} can now be formulated.
Let $A_m (\alpha^\prime,\alpha)$ be the acattering amplitude
corresponding
to $D_m$ and $\Gamma u=u$ on
$S_m:=\partial D_{m}, m=1,2$.

\begin{theorem}
If (2.18) and (3.17) hold then
\begin{equation}
        \rho(\delta) \leq c_1 \left(\frac{\ln |\ln \delta|}{|\ln \delta|}
        \right)^{c_2}, \quad
        \delta \to 0,
        \end{equation}
where $c_1, c_2$ are positive constants independent of $\delta >0$.
\end{theorem}

\begin{proof}
Let us sketch the steps of the proof.

\underline{Step 1.}
\begin{equation}
        \rho(\delta) \to 0 \hbox{\ as\ } \delta \to 0.
        \end{equation}
This follows from the uniquness theorem 3.2 and from the compactness of the
set $\cal O^\lambda$ in the space $C^{2, \mu}$, $\mu < \lambda$. For
simplicity of the
presentation we assume that $j=1$, that is, there is just one patch in the
covering of $S$, for example, $S$ is star-shaped.

\underline{Step 2.}
There exists an integer $m$ such that
\begin{equation}
        cd^m(x) \leq |w(x)| \leq c \varepsilon^{cd^c(x)}, \quad
        w(x) := u_1(x, \alpha) - u_2(x, \alpha),
        \end{equation}
where $d(x)$ is the distance from a point $x \in D^\prime_{12}$ to
$S_{12}$, we assume that $d(x) \sim \rho$,
$\varepsilon = ce^{-\gamma N(\delta)}$,
$\gamma = \const >0$, $N(\delta) = \frac{|\ln \delta|}{\ln|\ln \delta|}$,
and $c>0$ here and below stand for {\it various} constants indepent of
$\delta$ and $x$, and we assume that $d(x) \sim \rho$.
Symbol $d(x) \sim \rho$ means that
$c_1 d(x) \leq \rho \leq c_2 d(x)$ with some constants $c_1, c_2 >0$.

Let us show that (3.21) implies (3.19). From (3.21) one gets, replacing $d$
by $\rho$, dropping $w$ and taking $\log$:
\begin{equation}
        \ln \rho \leq c + c \rho^c \ln \varepsilon.
        \end{equation}
Recall that $c$ stands for different constants.

From (3.22) one gets
\begin{equation}
        \frac{\rho^c}{\ln \frac{1}{\rho}} \leq
        \frac{c}{\ln \frac{1}{\varepsilon}}.
        \end{equation}
Since $\rho^\beta < \frac{1}{\ln \frac{1}{\rho}}$ as $\rho \to 0$
and $\beta >0$, estimate
(3.23) implies
$$\rho \leq c \left(\frac{1}{\ln \frac{1}{\varepsilon}}\right)^
        {\frac{1}{c}} = c_1
        \left(\frac{\ln|\ln\delta|}{|\ln \delta|} \right)^{c_2},$$
where we have used the definition of $\varepsilon$,
namely $\varepsilon = ce^{-\gamma N(\delta)}$, which implies
$\frac{1}{\ln \frac{1}{\varepsilon}} \sim
\frac{\ln|\ln\delta|}{|\ln \delta|}$.

Let us give the details of the proof.

\underline{Step 1.}
Assume that (3.20) is false. Then $\rho(\delta) \geq c >0$
as $\delta \to 0$. Since $\cal O^\lambda$ is a compact set in
$C^{2, \mu}$, $\mu < \lambda$, one can
select sequences $S_{1n(\delta)}$ and
$S_{2n(\delta)}$ which converge in $C^{2, \mu}$ to $S_1$ and $S_2$
correspondingly, as $\delta \to 0$. Since $A(\alpha^\prime, \alpha)$
depends continuously on $S$ (see \cite{R20}, \cite{R7}) in the sense
\begin{equation}
        \lim_{n \to \infty} \sup_{\alpha^\prime, \alpha\in S^2}
        |A_n(\alpha^\prime, \alpha) - A(\alpha^\prime, \alpha)| = 0,
        \end{equation}
where the limit is taken in the process $S_n \to S$ in $C^{2, \mu}$,
$0 < \mu < \lambda$, one concludes that
$A_1(\alpha^\prime, \alpha) = A_2(\alpha^\prime, \alpha)$ for the
limiting surfaces $S_1$ and $S_2$. By the uniqueness theorem 3.1 it follows
that $S_1 = S_2$. Therefore $\rho := \rho(S_1, S_2) = 0$. However,
$\rho = \lim_{\delta \to 0} \rho (\delta) \geq c >0$.
This is a contradiction which proves (3.20).

\underline{Step 2.}
The function $w(x) := u_1(x, \alpha) - u_2(x, \alpha)$, where
$u_j (x, \alpha)$ is the scattering solution corresponding to the obstacle
$D_j$, $j =1,2,$ solves the equation
\begin{equation}
        (\nabla^2 + 1) w=0 \hbox{\ in\ } D^\prime_{12}, \quad
        D_{12} := D_1 \cup D_2,
        \end{equation}
satifies the radiation condition
\begin{equation}
        \lim_{r \to \infty} \int_{|s|=r}
        \left| \frac{\partial w}{\partial |x|} - i w \right|^2 ds = 0,
        \end{equation}
and
\begin{equation}
        \Vert w \Vert_{C^{2, \lambda}(D^\prime_{12})} \leq c.
        \end{equation}
It is proved in \cite{H},vol.3, p.14,
that solutions to elliptic second order
equations with smooth coefficients cannot have zeros of infinite order up
to the boundary without vanishing identically. This implies existence of an
integer $m>0$ for which the left inequality (3.21) holds.

The proof of the right inequality requires some preparations.

Let us sketch the steps of this proof.

\underline{Step 3.}
By the result of Theorem 2.5 one gets estimate (2.58)
in the region $|x|>a$, $B_a \supset D$.

Although in Theorem 2.5 the functions $u_1$ and $u_2$ were the solutions to
the Schr\"odinger equations with potentials vanishing in $B^\prime_a$, the
estimate (2.58) is proved for any solutions to equation (3.25) whose
difference satisfy (3.26) and (2.18). In particular, estimate (2.58) holds for
our $w$. Let us define $\varepsilon = c \mu(\delta) = ce^{-\gamma N(\delta)}$,
$N(\delta) = \frac{|\ln \delta|}{\ln |\ln \delta|}$.

\underline{Step 4.}
Let us prove the right inequality (3.21). Extend $w$ from
$D_{12}^\prime$ into $D_{12}$ so that the estimate similar to (3.27) holds:
\begin{equation}
        \Vert w \Vert_{C^{2, \lambda}(\R^3)} \leq c_1.
        \end{equation}
This is possible: using, for example, the known Stein's theorem one can
extend $u_j$ into $D_j$ since $S_j$ is $C^{2, \lambda}$ - smooth,
$j=1,2,$ and then $w:=u_1-u_2$ will be the $C^{2, \lambda}$- smooth extension
of $w$ from $D^\prime_{12}$ into $D_{12}$. Define
\begin{equation}
        f(x) := -(\nabla^2 + 1) w \hbox{\ in\ } \R^3,
        \end{equation}
where we denoted by $w$ the extended to $\R^3$ function $w(x)$. Then $f(x) =0$
in $D^\prime_{12}$, $f(x) \in C^\lambda (\R^3)$, and
\begin{equation}
        w(x) = \int_{D_{12}} \frac{e^{i|x-y|}}{4\pi|x-y|}
        f(y)dy, \quad x \in \R^3.
        \end{equation}
Denote $|x| = r$, $|y| = \rho$. Set $z:= re^{i\varphi}$. Choose any point
$x \in D^\prime_{12}$, in a neighborhood of $\widetilde{D_1}$, a connected
component of $D_{12} \backslash D_2$, such that
$d(x) := dist (x, S_1) \sim \rho := \rho(\delta)$. By step 1,
$\rho(\delta) \to 0$ as $\delta \to 0$, so that $\rho (\delta)$ is small
for small $\delta$.

Consider analytic continuation of $w(x)$, defined by (3.30), on the complex
$z$-plane (as in [11]).
Let $\omega$ be the angle between vectors $x$ and
$y$, $x \in D^\prime_{12}$. Since
$$ |x-y| = \sqrt{(r - \rho e^{i\omega})(r- \rho e^{-i\omega})},$$
this analytic continuation, that is, replacement $r\to z$,
in the expression  $|x-y|,$ 
is possible if $re^{i \varphi} - \rho e^{i\omega} \neq 0$ and
$re^{i \varphi} - \rho e^{-i\omega} \neq 0$. Choose a point $O$ on $S_1$
closest to $x$ and the coordinate system in which the origin is at $O$, and
the $x_1x_2-$ plane is tangent to $S_1$ at the point $O$. Since
$S_1$ is
sufficiently smooth, there exists a cone $K$ with an opening $\theta_0 >0$
and vertex at $O$ which belongs to $D^\prime_{12}$ and its axis passes
through the point $x$. The function $|x-y|$ admits analytic continuation in
the $z$-plane from the ray $r>0$ to the sector $|\varphi|< \theta_0$. Since
there are no points of $D_{12}$ inside the cone $K$, the expression
$$\zeta := \left[ (re^{i \varphi} -
  \rho e^{i\omega})(re^{i \varphi} - \rho e^{-i\omega})\right]^{\frac{1}{2}},
  \quad Im \zeta \geq 0, \quad r >0, \quad |\arg z |=|\varphi| < \theta_0,
  $$
does not vanish in the region $r>0$, $|\arg z|:=|\varphi|<\theta_0$.
Therefore the function (3.30), which we
denote $W(z)$, considered as a function of $z$, admits analytic continuation
in the sector $|\arg \varphi | < \theta_0, r>0$, and satisfies the following
inequalities there:
\begin{equation}
        |W(z)| \leq c, \quad |\arg z| < \theta_0, \quad r \geq 0,
        \end{equation}
\begin{equation}
        |W(z)| \leq \varepsilon, \quad \arg z = 0, \quad
        r \geq a,
        \end{equation}
where
\begin{equation}
        \varepsilon = ce^{-\gamma \frac{|\ln\delta|}{\ln|\ln \delta|}}.
        \end{equation}

One can map the sector $|\arg z| < \theta_0$, conformally onto the half-plane,
$z \to t = z^{\frac{\pi}{2\theta_0}}$, $|\arg t| <  \frac{\pi}{2}.$ Then
$W(z) := v(t)$, where $v(t)$ is analytic in the half-plane
$|\arg t| < \frac{\pi}{2}$, and satisfies there the inequalities
\begin{equation}
        |v(t)| \leq c, \quad |\arg t| < \frac{\pi}{2}, \quad
        |v(t)| < \varepsilon, \quad
        t \geq a_1.
        \end{equation}
The known two-constants- theorem (see \cite{EV}) and (3.34) imply:
\begin{equation}
        |v(t)| \leq c_1 \varepsilon^{h(t)},
        \end{equation}
where $h(t)$ is the harmonic measure corresponding to the domain $R$ on the
complex plane $t$ with the boundary consisting of the lines
$\{Re t = 0, \ -\infty < Im t < \infty\}$ and
$\{Im t = 0, Re t > a_1 > 0 \}$. Recall that $h=h(t_1, t_2)$,
 the harmonic measure, is a harmonic
function which solves the problem:
\begin{equation}
        \Delta h := h_{t_1t_1} + h_{t_2t_2} = 0
        \hbox{\ in\ } R,
        \end{equation}

$$h=0 \hbox{\ at\ } t_1 = 0, \quad
         -\infty < t_2 < \infty, \quad h=1
        \hbox{\ at\ }
        t_2 = 0, \quad t_1 \geq a,$$
$h(t_1, t_2)$ is bounded at infinity, $t = t_1 + it_2$.

By the maximum principle, $1>h(t_1, t_2)>0$ in $R$, and, by the Hopf lemma,
(see \cite{GT} p.34),
\begin{equation}
        \frac{\partial h}{\partial t_1} > 0 \hbox{\ at\ }
        t_1=0.
        \end{equation}
Thus
\begin{equation}
        h(t_1, 0) \geq ct_1, \quad
        0 < t_1 \leq t_0, \quad
        c=\const > 0,
        \end{equation}
where $t_0 >0$ is a sufficiently small number.

From (3.35) and (3.38) it follows that in a sufficiently small neighborhood
of the origin one has
$|v(t_1, 0)| \leq c \varepsilon^{ct_1}$. Returning to the $z$-variable one
gets
\begin{equation}
        |W(z)| \leq c\varepsilon^{c|z|^c}, \quad
        z = r = |x| > 0,
        \end{equation}
where $c>0$ stands for various constants.

Since $W(r) = w(x)$, inequality (3.39) is identical to the right inequaltiy
(3.21).

Theorem 3.2 is proved.
\end{proof}

\begin{remark}
An interesting open problem is to construct $S$ in the inverse obstacle
scattering problem analytically from noisy data
$A_\delta(\alpha^\prime, \alpha)$ in the way it is done in section (2.4),
formula (2.45), for the reconstruction of the potential in the
inverse potential scattering problem,
or even from exact data in the way it was done in section 2.2, formula (2.8).
In the next
section we prove that such a reconstruction formula does exist for
exact data for the inverse obstacle scattering problem.
\end{remark}

%%%%%%%%%%%%%%%%%%%%%%%%%%%%%%%%%%%%%%
\subsection{Existence of a reconstruction formula.}
Let $A(\alpha^\prime, \alpha)$ be the scattering amplitude corresponding
to an obstacle $D$, and assume, for example, that $\Gamma u=u$, so that
the Dirichlet condition holds on $S$. Take
$\alpha^\prime = \theta^\prime \in M$ in (3.6), multiply (3.6) by a
$\nu(\alpha, \theta)$ and integrate with respect to $\alpha$ over
$S^2$ to get:
\begin{equation}
        -4\pi \int_{S^2} A(\theta^\prime, \alpha) \nu (\alpha, \theta)
        d \alpha = \int_{S} e^{-i \theta^\prime \cdot s}
        \int_{S^2} u_N(s, \alpha) \nu (\alpha, \theta) d \alpha\,  ds.
        \end{equation}

Here $\nu(\alpha,\theta)\in L^2(S^2)$ is some function which is
chosen so that (3.41) holds.
We prove below the following lemma:

\begin{lemma}
The set $\{u_N(s, \alpha)\}_{\forall \alpha \in S^2}$ is total in
$L^2(S)$.
\end{lemma}
This implies that, given an arbitrary small number $\eta >0$ and
$\theta \in M$, there
exists a $\nu_\eta (\alpha, \theta) \in L^2(S^2)$ such that
\begin{equation}
        \Vert \int_{S^2} u_N (s, \alpha) \nu_\eta (\alpha, \theta)
        d \alpha - \frac{\partial e^{i \theta \cdot s}}{\partial N}
        \Vert_{L^2(S)} \leq \eta.
        \end{equation}
From (3.40), with $\nu = \nu_\eta$, and (3.41) one gets:
\begin{equation}
        -4\pi \lim_{\eta \to 0} \int_{S^2} A(\theta^\prime, \alpha) \nu_\eta
        (\alpha, \theta) d\alpha = \int_S e^{-i \theta^\prime \cdot s}
        \frac{\partial e^{i \theta \cdot s}}{\partial N} ds
        \end{equation}
Let us assume $\theta, \theta^\prime \in M$, $\theta - \theta^\prime = \xi$,
$\xi \in \R^3$ is an arbitrary fixed vector.

By Green's formula one gets:
\begin{equation}
        \int_S e^{-i \theta^\prime \cdot s}
        \frac{\partial e^{i \theta\cdot s}}{\partial N} ds
        = \frac{1}{2} \int_S
        \frac{\partial e^{i(\theta - \theta^\prime)\cdot s}}{\partial N}
        ds = -\frac{|\xi|^2}{2} \int_D e^{i \xi \cdot x} dx
        = - \frac{|\xi|^2}{2} \widetilde{\chi_D} (\xi),
        \end{equation}
where $\chi_D(x) = 1$ in $D$, $\chi_D(x) = 0$ in $D^\prime$.

From (3.42) and (3.43) one gets an {\it inversion formula for finding $D$
from the data $A(\alpha^\prime, \alpha)$:}
\begin{equation}
        \widetilde{\chi_D} (\xi) = 8\pi |\xi|^{-2} \lim_{\eta \to 0}
        \int_{S^2} A(\theta^\prime, \alpha)
        \nu_\eta (\alpha, \theta) d \alpha.
        \end{equation}

Before proving the basic Lemma 3.1, a remark is in order: in contrast to
our theory for inverse potential scattering problem (see the inversion
formula (2.8) and the algorithm for calculating the function
$\nu(\alpha, \theta)$ in (2.15)) we do not give an algorithm for
calculating the function $\nu_\eta(\alpha, \theta)$ in (3.44).

{\it Finding such an algorithm is an open problem.}

{\it We now prove Lemma 3.1}: assume the contrary. Then there exists a function
$f \in L^2(S)$ such that
\begin{equation}
        \int_S f(s) u_N (s, \alpha) ds = 0 \quad
        \forall \alpha \in S^2.
        \end{equation}
Define
\begin{equation}
        v(y) := \int_S f(s) G_N(s, y) ds,
        \end{equation}
where $G(x,y)$ is the Green function:
\begin{equation}
        (\nabla^2 + 1) G(x,y) = -\delta(x-y) \hbox{\ in\ } D^\prime,
        \end{equation}
\begin{equation}
        G(x,y) = 0, \quad x \in S,
        \end{equation}
$G$ satisfies the radiation condition (1.6).

We claim that (3.45) implies
\begin{equation}
        v(y) = 0 \hbox{\ in\ } D^\prime,
        \end{equation}
where $v(y)$ is defined in (3.46).

Indeed,
\begin{equation}
        (\nabla^2 + 1) v = 0 \hbox{\ in\ } D^\prime,
        \end{equation}
and
\begin{equation}
        v(y) = o\left(\frac{1}{|y|} \right) \hbox{\ as\ } |y| \to \infty.
        \end{equation}
The relation (3.51) follows from (3.45) and (1.7).

It is well known \cite{R7}, p. 25, that (3.50) and (3.51) imply (3.49).

If (3.49) holds, then, taking $y \to\sigma\in S$
(along the normal to $S$ at the point $\sigma$) in (3.46), one gets
\begin{equation}
        f(s) = 0.
        \end{equation}
Indeed, if $\sigma \in S$ then
\begin{equation}
        G_N(s,y) \to \delta_S (s- \sigma) \hbox{\ as\ }
        y \to \sigma,
        \end{equation}
where $\delta_S (s- \sigma)$ is the delta-function on the surface
$S$ (see formula (3.58) below).

From (3.53), and (3.49) and (3.46) formula (3.52) follows.

Let us give a proof of (3.53).

Consider the problem:
\begin{equation}
        (\nabla^2 + 1) v = 0 \hbox{\ in\ } D^\prime,
        \end{equation}
\begin{equation}
        v=f (s) \hbox{\ on\ } S,
        \end{equation}

\begin{equation}
v \hbox{\ satisfies\ } \hbox{\ the\ } \hbox{\ radiation\ }
\hbox{\ condition\ } (3.26).
\end{equation}

This problem has a unique solution representable by the Green formula:
\begin{equation}
        v(x) = \int_S f(s) G_N (s,x)ds.
        \end{equation}

Since (3.55) holds for this unique solution (3.57) taking
$x \to \sigma \in S$ one gets (3.53), where $\delta_S (s-\sigma)$ is the
distribution which acts by the formula
\begin{equation}
        \int_S f(s) \delta_S(s-\sigma) ds = f(\sigma).
        \end{equation}

This argument requires $f(s)$ to be continuous if one understands the
delta-function in the usual sense. However, if $\delta_S(s-\sigma)$
is understood as the kernel (in the distributional sense) of the identity
operator in some space of functions $f(s)$, for which problem (3.54)-(3.56)
has a unique solution, then (3.58) makes sense in the space for example,
in $L^2(S)$, and (3.58) is understood in this case as equality of the
elements of this space. In the case of $L^2(S)$ this means that the
equality holds almost everywhere on $S$ with respect to two-dimensional
Hausdorff  measure on $S$. Lemma 3.1 is proved.

Thus, formula (3.44) is proved. \qed

%%%%%%%%%%%%%%%%%%%%%%%%%%%%%%%%%%%%%%%%%
\section{Limiting procedure and stability estimates}%4

Let $D \subset \R^3$ be a bounded domain, $\chi (x)$ be the characteristic
function of $D$, $t >0$ be a parameter, $q(x) = t \chi(x)$ be the potential.
Consider the potential scattering
problem (1.1)--(1.2). We prove that the scattering
solution $u(x, \alpha; t)$ converges,
as $t \to +\infty$, to $u(x, \alpha)$, the scattering solution
corresponding to the obstacle $D$, and 
$u(x,\alpha)$ satisfies
the Dirichlet boundary condition on
$S := \partial D$. This result is old \cite{R22}. We also prove the following,
more recent estimates \cite{R13}:
\begin{equation}
        \Vert u(x, \alpha; t) \Vert_{L^2(D)} \leq
        \frac{c}{t^{\frac{1}{2}}}, \quad
        \Vert u (x, \alpha; t) - u(x,\alpha)
        \Vert_{L^2(\widetilde{D}^\prime)} \leq
        \frac{c}{t^{\frac{1}{2}}},
        \end{equation}
\begin{equation}
        \Vert \nabla u(x, \alpha; t) \Vert_{L^2(B_R)} \leq
        c(R), \quad B_R \supset D,
        \end{equation}
\begin{equation}
        \Vert u(x, \alpha; t)\Vert_{L^2(S)} \leq
        \frac{c}{t^{\frac{1}{2}}},
        \end{equation}
where $\widetilde{D}^\prime$ is a strictly inner compact subdomain of
$D^\prime := \R^3 \backslash D$.

Assume there are two obstacles $D_1$ and $D_2$,
$D_j \subset {\mathcal O}^{2, \lambda}$, $0 < \lambda \leq 1$.
Let
$$ A(\alpha^\prime, \alpha; t_1, t_2):= A_1(\alpha^\prime, \alpha; t_1)-
   A_2(\alpha^\prime, \alpha; t_2).
   $$

Then we prove the following stability estimate:
\begin{equation}
        \sup_{\alpha^\prime, \alpha \in S^2}
        \left|A(\alpha^\prime, \alpha; t_1, t_2) \right| \leq
        c\left(\frac{1}{t_{12}^{\frac{1}{2}}} + \rho \right),
        \end{equation}
where $t_{12} = \min (t_1, t_2)$ and $\rho$ is the symmetric Hausdorff
distance between $S_1$ and $S_2$, defined by (3.18).

Let us prove the above estimates.

One has
\begin{equation}
        \left[ \nabla^2 + 1- t\chi (x) \right] u(x, \alpha; t) = 0
        \hbox{\ in\ } \R^3.
        \end{equation}
Define
\begin{equation}
        \Vert u \Vert^2 := \int_{\R^3}
        \frac{|u (x)|^2 dx}{(1+ |x|^2)^\frac{\sigma}{2}}, \quad
        \sigma > 1.
        \end{equation}
We drop the $\alpha$-dependence in $u(x, \alpha; t)$.

Assume that
\begin{equation}
        \Vert u(x;t)\Vert \leq c,
        \end{equation}
where $c = \const >0$ does not depend on $t$.

As $t \to +\infty$, one can select, using (4.7), a weakly convergent in the
norm (4.6) sequence, denoted again $u$. Thus
\begin{equation}
        u(x, t) \rightharpoonup u.
        \end{equation}
By elliptic estimates (see formula (6.1) below), formulas (4.8) and
(4.5) imply
\begin{equation}
        \Vert u(x; t) \Vert_{H^2(\widetilde{D}^\prime)} \leq c,
        \quad \Vert u(x, t) - u(x) \Vert_{H^2(\widetilde{D}^\prime)}
\to 0 \hbox{\ as\ } 
        t \to \infty.  
        \end{equation}
Multiply (4.5) by $\overline{u} (x, t)$ and integrate over a ball
$B_R \supset D$, to get
\begin{equation}
        \int_{B_R} \left[|\nabla u(x; t)|^2 + t|u(x;t)|^2 \right] dx =
        \int_{B_R} |u(x;t)|^2 dx + \int_{\partial B_R}
        \frac{\partial u}{\partial N} \overline{u}\, ds
        \end{equation}
where the bar stands for complex conjugate.

From (4.10), (4.7) and (4.9) one gets
\begin{equation}
        \int_{B_R} |\nabla u(x;t)|^2 dx \leq c(R), \quad
        \int_{B_R}|u(x;t)|^2 dt \leq
        \frac{c(R)}{t}.
        \end{equation}
This yields (4.2) and the first inequality in (4.1).

Let us prove (4.3). The embedding theorem yields (see, e.g.,\cite{M},
p. 66):
\begin{equation}
        \Vert u(x; t) \Vert_{L^2(S)} \leq c
        \left(\varepsilon \Vert \nabla u (x;t)\Vert_{L^2(D)}
          + \varepsilon^{-1}
        \Vert u (x; t)\Vert_{L^2(D)} \right),
        \quad 0\leq \varepsilon <1.
        \end{equation}
Take $\varepsilon = t^{-\frac{1}{2}}$ and use (4.11). Then (4.12) 
yields (4.3). Let us prove the second inequality (4.1).

Denote $v:= u(x;t) - u(x)$. Then
\begin{equation}
        \nabla^2 v+v = 0 \hbox{\ in\ } D^\prime; \quad
        v \hbox{\ satisfies\ } (3.26),
        \end{equation}
\begin{equation}
        v = u(x;t) \hbox{\ on\ } S.
        \end{equation}
By Green's formula, one gets
\begin{equation}
        v(x) = \int_S u(s; t)
        \frac{\partial G(s,x)}{\partial N_s} ds,
        \end{equation}
where $G(x,y)$ is the Green function for the Dirichlet operator
$\nabla^2 + 1$ in $D^\prime$.
\begin{equation}
        |v(x)| \leq \Vert u(s;t) \Vert_{L^2(S)}
        \Vert \frac{\partial G(s,x)}{\partial N_s} \Vert_{L^2(S)}.
        \end{equation}
From (4.16) and (4.3) the second estimate (4.1) follows.

It is clear that the function $u=u(x)$ in (4.8) is the scattering solution
$u(x, \alpha)$ corresponding to the obstacle scattering problem with the
Dirichlet condition on $S$. Indeed, $u=0$ on $S$ (see(4.3)), $u$ solves
equation (4.13) in $D^\prime$, $u-e^{i \alpha \cdot x}$ satisfies the
radiation condition. These three conditions determine $u$ uniquely, and since
$u(x, \alpha)$ satisfies these conditions, it follows that $u(x, \alpha)$.

Let us prove (4.4).

If $D_1 = D_2 = D$, then formula (1.30) yields
\begin{equation}
        \left|A(\alpha^\prime, \alpha; t_1, t_2) \right| \leq
        \left|t_2-t_1\right|
        \Vert u_1(x, \alpha; t_1)\Vert_{L^2(D)}
        \Vert u_2(x, -\alpha^\prime; t_2)\Vert_{
        L^2(D)}\leq
        c\frac{|t_2-t_1|}{(t_1 t_2)^{\frac{1}{2}}}.
        \end{equation}
If $t_1 = t_2 = +\infty$, then formula (3.4) yields:
\begin{equation}
        |A(\alpha^\prime, \alpha)| \leq c \rho(S_1, S_2).
        \end{equation}
In the general case a combination of the above estimates yields (4.4)
(see \cite{R13}) and also the estimate
\begin{equation}
        |A(\alpha^\prime, \alpha; t_1, t_2)| \leq c \left[|t_1-t_2| +
        \rho(S_1, S_2)\right]
        \end{equation}
useful when $|t_1-t_2|$ is small and $t_1, t_2 \in [1, t_0]$, where
$t_0 >1$ is fixed.

If $D_1 = D_2$, then (4.19) yields
\begin{equation}
        |A(\alpha^\prime, \alpha; t_1, t_2)| \leq c|t_1-t_2|.
        \end{equation}

%%%%%%%%%%%%%%%%%%%%%%%%%%%%%%%%%%%%%%%%%
\section{Inverse geophysical scattering with fixed-frequency data}%5
Consider the problem
\begin{equation}
        \left[\nabla^2 +1+v(x)\right] w= -\delta(x-y) \hbox{\ in\ }
        \R^3,
        \end{equation}
\begin{equation}
        \lim_{r \to \infty} \int_{|S|=r}
        \left|\frac{\partial w}{\partial |x|} - iw \right|^2 ds =0,
        \end{equation}
where $v(x) \in L^2(\R^3)$ is a compactly supported real-valued function with
support in $\R^3_-$, the lower half-space. In acoustics $u$ has the physical
meaning of the pressure, $v(x)$ is the inhomogeneity in the velocity profile.
We took the fixed wavenumber $k=1$ without loss of generality. The source
$y$ is on the plane $P := \{x :x_3 = 0\}$, i.e., on the surface of the
Earth, the receiver $x \in P$.

The data are the values $\{w(x,y)\}_{\forall x,y \in P}$.

The inverse geophysical scattering problem is: {\it given the above data,
find $v(x)$.}

The uniqueness theorem for the solution to this problem is obtained in
\cite{R23}, \cite{R2}.

Problem (5.1) -(5.2) differs from the inverse potential scattering by the
source: it is a point source in (5.1) and a plane wave in (1.2). Let us show
how to reduce the inverse geophysical scattering problem to inverse
potential scattering problem using the ``lifting"
\cite{R2}, \cite{R28}.
Suppose the data
$w(x,y)$, $x \in P$, $y \in P$, are given. Fix $y$ and solve the problem:
\begin{equation}
        (\nabla^2 + 1) w = 0 \hbox{\ in\ } \R^3_+ = \{x : x_3 >0\},
        \end{equation}
\begin{equation}
        w = w(x,y), \quad x \in P,
        \end{equation}
\begin{equation}
        w \hbox{\ satifies\ } (5.2).
        \end{equation}

This problem has a unique solution and there is a Poisson-type analytical
formula for the solution to (5.3)-(5.5), since the Green function of the
Dirichlet operator $\nabla^2 +1$ in the half-space $\R^3_+$ is known
explicitly, analytically:
\begin{equation}
        G_1(x,y) = \frac{e^{i|x-y|}}{4\pi|x-y|} -
        \frac{e^{ik|x-\overline{y}|}}{4\pi|x-\overline{y}|}, \quad
        \overline{y} := (y_1, y_2, -y_3).
        \end{equation}

Therefore the data $w(x,y) \forall x \in P$ determine uniquely and explicitly
(analytically) the data $w(x,y) \forall x \in \R^3_+$, $y \in P$.
We have lifted the data from $P$ to $\R^3_+$ as far as $x$-dependence is
concerend and get $w(x,y) \forall x, y \in \R^3_+$ given
$w(x,y) \forall x,y \in P$.

If $w(x,y)$ is known for all $x,y \in \R^3_+$, then one uses formula (1.7)
and calculates analytically the scattering solution $u(x, \alpha)$
corresponding to the potential $q(x) := -v(x)$ and $k=1$, where
$\alpha \in S^2_- := \{\alpha: \alpha \in S^2, \alpha_3 \leq 0\}$.
Given $u(x, \alpha)$ for all $x \in \R^3_+$ and $\alpha \in S^2_-$, one
can calculate the scattering amplitude $A(\alpha^\prime, \alpha)$
$\forall \alpha^\prime \in S^2_+ :=
\{\alpha: \alpha \in S^2, \alpha_3 \geq 0\}$.

If the scattering amplitude $A(\alpha^\prime, \alpha)$, corresponding to
the compactly supported $q(x) =-v(x) \in L^2(\R^3)$ is known
$\forall \alpha^\prime \in S^2_+, \forall \alpha \in S^2_-$, then the
uniqueness
of the solution to inverse geophysical problem follows from theorem 2.1.

Stability estimates obtained for the solution to inverse potential
scattering problem with fixed-energy data remain valid for the inverse
geophysical problem: via the lifting process one gets the scattering
amplitude $A(\alpha^\prime, \alpha)$ corresponding to the potential
$q(x) = -v(x)$, and the stability estimates for $\widetilde{q}(\xi)$,
obtained in sections 2.3 and 2.4, yield stability estimates for
$\widetilde{v}(\xi) = -\widetilde{q}(\xi)$.

Practically, however, there are two points to have in mind. The first point
is: if the noisy data $u_\delta(x,y)$ are given, where
$\sup_{x,y \in P} |u_\delta(x,y) - u(x,y)| < \delta$, then one has to
overcome the following difficuly in the lifting process: data
$\varphi (x,y), x, y \in P$, such that $|\varphi(x,y)|< \delta$, may not
decay as $|x| \to \infty$, $|y| \to \infty$ on $P$, and this brings the main
difficulty.

The second point is: if one uses the inversion algorithms presented in
sections 2.2-2.4, then one uses the data
$A(\alpha^\prime, \alpha) \forall \alpha^\prime, \alpha \in S^2$. Of course,
the exact data
$A(\alpha^\prime, \alpha) \forall \alpha^\prime \in S^2_+,\alpha \in S^2$,
determine uniquely the data $A(\alpha^\prime, \alpha)$
$\forall \alpha^\prime, \alpha \in S^2$,
but practically finding the full data from the partial data is an
ill-posed problem.

%%%%%%%%%%%%%%%%%%%%%%%%%%%%%%%%%%%%%%%%%
\section{Proofs of some estimates.}%6
Here we prove some technical results used above: estimates 
(1.18), (1.19),
(1.20), (1.30), (2.13), and (2.17).

%%%%%%%%%%%%%%%%%%%%%%%%%%%%%%%%%%%%%%
\subsection{Proof of (1.18).} 
It is sufficient to prove (1.18) with $L^2(D)$ in place of $H^2(D)$:
since $w(x)$ and
$$\int_{S^2} u(x, \alpha) \nu_\varepsilon (\alpha) d \alpha$$
solve equation (1.1), the elliptic estimate (see \cite{GT})
\begin{equation}
        \Vert \varphi \Vert_{H^2(D_1)} \leq c
        \left[ \Vert L \varphi \Vert_{L^2(D_2)} +
        \Vert \varphi \Vert_{L^2(D_2)} \right], \quad
        D_1 \subset D_2,
        \end{equation}
where $L= \nabla^2 + 1-q(x)$, $D_1$ is strictly inner subdomain of
$D_2$ and $c = c(D_1, D_2) = \const >0$, implies that
$\Vert \varphi \Vert_{H^2(D_1)} \leq c
\Vert \varphi \Vert_{L^2(D_2)}$ if $L \varphi = 0$. If (1.18), with
$k=1$ and $L^2(D)$ in place of $H^2(D)$, is false then
\begin{equation}
        \int_D w(x) \int_{S^2} u(x, \alpha) \nu(\alpha) d \alpha = 0
        \quad \forall \nu(\alpha) \in L^2(S^2).
        \end{equation}
Therefore
\begin{equation}
        \int_D w(x) u(x, \alpha) dx = 0 \quad
        \forall \alpha \in S^2.
        \end{equation}
This implies
\begin{equation}
        \int_D w(x) G(x,y) dx =0 \quad
        \forall y \in D^\prime,
        \end{equation}
where $G(x,y)$ is the Green function of the operator $L$.

Indeed, denote the integral on the left-hand side of (6.4) by
$\varphi(y)$. Then
\begin{equation}
        L \varphi = 0 \hbox{\ in\ } D^\prime; \quad
        \varphi = o\left(\frac{1}{|y|}\right) \hbox{\ as\ }
        |y| \to \infty.
        \end{equation}
The second relation (6.5) follows from (6.3) and (1.7).
From (6.5) one gets
(6.4) by lemma 1 on p.25 in \cite{R7}. From (6.4) it follows that
\begin{equation}
        L \varphi = -w(x) \hbox{\ in\ } D, \quad
        \varphi = 0 \hbox{\ in\ } D^\prime, \quad
        \varphi \in H^2_{loc} (\R^3).
        \end{equation}
Thus
\begin{equation}
        L \varphi = -w \hbox{\ in\ } D,\,\, \varphi = \varphi_N = 0
\hbox{\ on\ } S.
        \end{equation}
Multiply (6.7) by $\overline{w}$, integrate over $D$, then by parts using
the boundary conditions (6.7), use the equation $Lw=0$ and get
\begin{equation}
        \int_D |w(x)|^2 dx = 0.
        \end{equation}
Thus $w(x) = 0$. Estimate (1.18) is proved. \qed

%%%%%%%%%%%%%%%%%%%%%%%%%%%%%%%%%%%%%%
\subsection{Proof of (1.20) and (1.21).}  
From (1.19) one gets
\begin{equation}
        \nabla^2 R + 2i\theta \cdot \nabla R-q(x) R = q(x)
        \hbox{\ in\ } \R^3.
        \end{equation}
Denote $L = \nabla^2 + 2i\theta \cdot \nabla$, and define
\begin{equation}
        w(x) := L^{-1} f = \frac{1}{(2\pi)^3}
        \int_{\R^3}
        \frac{\widetilde{f}(\xi) e^{i \xi \cdot x}}{\xi^2 + 2\xi \cdot \theta}
        d \xi.
        \end{equation}
Note that $Lw=-f(x)$.
We will prove below that (see \cite{R2},  \cite{R6}):
\begin{equation}
        \Vert L^{-1} f\Vert_{L^\infty(D_1)} \leq c
        \left(\frac{\ln |\theta|}{|\theta|}\right)^{\frac{1}{2}}
        \Vert f \Vert_{L^2(D)}, \quad
        \theta \in M, \quad
        |\theta| \to \infty,
        \end{equation}
where $D_1$ is an arbitrary compact domain
$c=c(D_1, \Vert q \Vert_{L^2(B_a)} )$, $D \subset B_a$.

We will also prove that
\begin{equation}
        \Vert L^{-1} f \Vert_{L^2(D_1)} \leq
        \frac{c}{|\theta|} \Vert f \Vert_{L^2(D)}, \quad
        |\theta| \to \infty, \quad
        \theta \in M.
        \end{equation}

{\it Let us show that (6.11) implies existence of the
special solutions (1.19).}

If (6.11) and (6.12) hold, then (1.20) and (1.21) are easily derived.

Indeed, rewrite (6.9) as
\begin{equation}
        R = L^{-1} qR + L^{-1} q.
        \end{equation}
From (6.11) and (6.13) it follows that
\begin{equation}
        \Vert L^{-1}qR \Vert_{L^\infty(D)} \leq c
        \left(\frac{\log |\theta|}{|\theta|}\right)^{\frac{1}{2}}
        \Vert qR \Vert_{L^2(D)} \leq c
        \left(\frac{\log|\theta|}{|\theta|}\right)^{\frac{1}{2}}
        \Vert q \Vert_{L^2(D)}
        \Vert R \Vert_{L^\infty(D)}.
        \end{equation}
Therefore the operator $L^{-1}q : L^\infty(D) \to L^\infty(D)$ has the
norm going to zero as $|\theta| \to \infty$, $\theta \in M$. Thus
equation (6.13) is uniquely solvable in $L^\infty(D)$ if
$|\theta| \gg  1$, $\theta \in M$. Moreover, the following estimate holds:
\begin{equation}
        \Vert R \Vert_{L^\infty(D)} \leq c 
        \Vert L^{-1}q \Vert_{L^\infty(D)} \leq c
        \left(\frac{\ln|\theta|}{|\theta|}\right)^{\frac{1}{2}}
        \Vert q \Vert_{L^2(D)}.
        \end{equation}

Estimate (1.20) follows.

To derive (1.21) from (6.12) one writes
\begin{equation}
        \Vert L^{-1} q R\Vert_{L^2(D)} \leq \frac{c}{|\theta|}
        \Vert qR \Vert_{L^2(D)} \leq \frac{c}{|\theta|}
        \Vert q \Vert_{L^2(D)} \Vert R \Vert_{L^\infty(D)}.
        \end{equation}
Therefore (6.13), (6.15) and (6.16) yield (1.21):
\begin{equation}
        \Vert R \Vert_{L^2(D)} \leq \frac{c}{|\theta|}
        \Vert q \Vert^2_{L^2(D)}
        \left(\frac{\log |\theta|}{|\theta|}\right)^{\frac{1}{2}} +
        \Vert L^{-1}q \Vert_{L^2(D)} \leq \frac{c}{|\theta|}.
        \end{equation}

%\begin{proof}
{\it Proof of (6.11).}
If $\theta \in M$ then $\theta = a+ib$,
$a, b \in \R^3$, $a \cdot b = 0$, $a^2-b^2 = 1$. Choose the coordinate
system such that $a = \tau e_2$, $b=t e_1$,
$\tau = (1 + t^2)^{\frac{1}{2}}$, $e_j$, $1 \leq j \leq 3$, are
the orthonormal basis vectors. Then
\begin{equation}
        \xi^2 + 2\theta \cdot \xi = \xi^2_1 + \xi^2_2 + \xi^2_3 +
        2 \tau \xi_2 + 2it\xi_1 = \xi^2_1 +
        (\xi_2 + \tau)^2 + \xi^2_3 - \tau^2 + 2it\xi_1.
        \end{equation}

This function vanishes if and only if
\begin{equation}
        \xi_1 = 0, \quad (\xi_2 + \tau)^2 + \xi^2_3 = \tau^2.
        \end{equation}

Equation (6.19) defines a circle $C_{\tau}$ or radius $\tau$ in the plane
$\xi_1 = 0$ centered at $(0, -\tau, 0)$. Let $T_\delta$ be a
toroidal neighborhood of $C_{\tau}$, where the section of the
torus by a plane orthogonal to $C_{\tau}$ is a square with size $2\delta$
and the center at $C_{\tau}$.

Denote $u(x) := L^{-1}f$, where $L^{-1}f$ is defined in (6.10). One has
\begin{equation}
        \left| u(x)\right| \leq \frac{1}{(2\pi)^3}
        \left| \int_{T_\delta}
        \frac{\widetilde{f}(\xi) e^{i\xi \cdot x}d\xi}
        {\xi^2+2\xi \cdot \theta}\right| + \frac{1}{(2\pi)^3}
        \left| \int_{\R^3 T_\delta}
        \frac{\widetilde{f}(\xi)e^{i\xi \cdot x}d\xi}{\xi^2 +2\xi \cdot \theta}
        \right| := I_1 + I_2,
        \end{equation}
\begin{equation}
  \begin{split}
   I_1 &\leq c \Vert \widetilde{f} \Vert_{L^\infty(\R^3)}
        \int_{T_\delta} \frac{d\xi}{|\xi^2+2\xi \cdot \theta|} \\
   &\leq
     c\Vert f \Vert_{L^1(\R^3)} \int^\delta_{-\delta} d\xi_1
      \int^{2\pi}_0 d \varphi \int^{\tau+\delta}_{\tau - \delta}
      \frac{\rho d \rho}{\sqrt{4t^2\xi^2_1+(\xi^2_1 + \rho^2-\tau^2)^2}}\\
   &= c \Vert f \Vert_{L^1(\R^3)} \int^\delta_0
     \int^{xi_1^2+2\tau\delta+\delta^2}_{\xi^2_1-2\tau\delta+\delta^2}
     \frac{d\mu}{\sqrt{4t^2\xi^2_1+\mu^2}} \\
  &\leq c(D) \Vert f \Vert_{L^2(D)} \int^\delta_0 d\xi_1\int^{3\tau\delta}_0
     \frac{d \mu}{\sqrt{4t^2\xi^2_1+ \mu^2}},\ 0< \delta < \frac{\tau}{2},
        \end{split}
        \end{equation}
where $\rho^2 = (\xi_2+ \tau)^2 + \xi^2_3$ and we have used the
Cauchy inequality        
$\Vert f \Vert_{L^1(\R^3)} = \Vert f \Vert_{L^1(D)} \leq c(D)
\Vert f \Vert_{L^2(D)}$ and an elementary inequality
$\xi^2_1 + 2\tau \delta + \delta^2 \leq 3 \tau \delta$,
which holds if $\xi^2_1 \leq \delta^2$ and $\tau > 2\delta$.

Let $\beta := 2t\xi_1$. Then
\begin{equation}
        \frac{1}{2t} \int^{2t\delta}_0 d\beta \int^{3\tau\delta}_0
        \frac{d \mu}{\sqrt{\beta^2 +\mu^2}} \leq
        \frac{1}{2t} \int^{3(t+\tau)\delta}_0 d\rho\rho
        \int^{\frac{\pi}{2}}_0 d \varphi
        \frac{1}{\rho} = \frac{\pi}{4t}
        3(t + \tau)\delta \leq c\delta,
        \end{equation}
where we have used the relations $\frac{\tau}{t} \to 1$ as $t \to \infty$
and took into account that
$t\to \infty$ if $|\theta|\to\infty$.
From (6.21) and (6.22) one gets
\begin{equation}
        I_1 \leq c \Vert f \Vert_{L^2(D)} \delta.
        \end{equation}
By $c>0$ we denote various constants independent of $\delta$ and $t$.

Let us estimate $I_2$:
\begin{equation}
        I^2_2 \leq c \Vert \widetilde{f} \Vert_{L^2(\R^3)}
        \int_{\R^3 \backslash T_\delta}
        \frac{d\xi}{|\xi^2+2\theta \cdot \xi|^2} =
        c\Vert f \Vert^2_{L^2(D)} {\cal J},
        \end{equation}
where the Parseval equality was used and by ${\mathcal J}$
the integral in (6.24) is denoted. One has
\begin{equation}
        {\cal J} \leq \int_{|\xi_1|>\delta}
        \frac{d\xi}{|\xi^2+2\theta \cdot \xi|^2} +
        \int_{|\xi_1 < \delta, |\rho-\tau| \geq \delta}
        \frac{d\xi}{|\xi^2 + 2\theta \cdot \xi|^2} :=
        j_1+j_2.
        \end{equation}

{\it Let us estimate $j_1$}:
\begin{equation}
    \begin{split}
    j_1 &\leq c \int^\infty_\delta    d\xi_1
       \int^\infty_0
       \frac{\rho d \rho}{4\xi^2_1 t^2 + (\xi^2_1 + \rho^2 - \tau^2)^2}\\
    &\leq c \int^\infty_\delta 
        d\xi_1 \int^\infty_{\xi^2_1 -\tau^2}
         \frac{d\mu}{4\xi^2_1 t^2+\mu^2}
        \leq c \int^\infty_\delta \frac{d\xi_1}{\xi_1 t}
         \left(\frac{\pi}{2} - \arctg \frac{\xi^2_1-\tau^2}{2\xi_1 t}
        \right)
        \end{split}
        \end{equation}
Let $\frac{\xi_1}{2t}=x$. Then the integral on the right-hand side of (6.26)
can be written as:
\begin{equation}
        j_1 \leq \frac{c}{t} \int^\infty_{\frac{\delta}{2t}}
        \frac{dx}{x} \left[\frac{\pi}{2} -\arctg \left(x-\frac{\tau^2}{4t^2}
        \frac{1}{x}\right) \right].
        \end{equation}
If $t \to \infty$, then $\frac{\tau^2}{4t^2} \to \frac{1}{4}$.
Let us use the elementary inequalities:
\begin{equation}
        \frac{\pi}{2} - x \leq \arctg \frac{1}{x}, \quad
        0 < x \leq \frac{\pi}{2};
        \end{equation}
\begin{equation}
        \arctg \frac{1}{x} \leq \frac{\pi}{2} - \frac{x}{2}, \quad
        x \to +0.
        \end{equation}

Then
\begin{equation}
        \frac{\pi}{2} - \frac{1}{y} \leq\arctg y \leq \frac{\pi}{2} -
        \frac{1}{2y}, \quad
        y \to +\infty.
        \end{equation}
Thus, with $A := \frac{\tau^2}{4t^2}$, one gets
\begin{equation}
        \frac{1}{x} \left[ \frac{\pi}{2} -
        \arctg \left(x-\frac{A}{x}\right)\right] \leq \frac{1}{x}
        \left(x - \frac{A}{x}\right)^{-1} \leq
        \frac{c}{x^2}, \quad
        x \to +\infty,
        \end{equation}
and
\begin{equation}
        \frac{1}{x} \left[\frac{\pi}{2} - \arctg\left(x -\frac{A}{x}\right)
        \right] \leq \frac{c}{x}, \quad
        x \to +0.
        \end{equation}
From (6.3), (6.4) and (6.27) one gets
$$j_1 \leq \frac{c}{t}\left(\int^1_{\frac{\delta}{2t}} \frac{dx}{x} +
c \right) \leq \frac{c}{t} \ln \frac{\delta}{2t}, \quad t \to +\infty,$$
so
\begin{equation}
        j_1 \leq c \frac{\ln \frac{\delta}{2t}}{t}, \quad
        t \to +\infty.
        \end{equation}

{\it Let us estimate $j_2$}:
\begin{equation}
   \begin{split}
     j_2 = &\int^\delta_{-\delta} d\xi_1
        \int_{\substack{\rho > \tau + \delta \\ 0<\rho \leq \tau -\delta}}
        d\rho\rho \int^{2\pi}_0
        d\varphi
        \frac{1}{4\xi^2_1 t^2+(\xi_1^2+\rho^2-\xi^2)^2} \\
      & \leq   c \int^\delta_0 d\xi_1 
        \left\{ \int^\infty_{\xi^2_1+(\tau+\delta)^2-\tau^2}
        \frac{d\mu}{4\xi^2_1t^2+\mu^2} +
        \int^{\tau^2 -\xi_1^2}_{2\tau\delta -\delta^2-\xi^2_1}
        \frac{d\nu}{4\xi^2_1t^2 +\nu^2} \right\}
        \end{split}
        \end{equation}
        
where $\mu = \xi^2_1 + \rho^2-\tau^2$ and $\nu = \tau^2-\rho^2-\xi^2_1$.

One has:
\begin{equation}
        \begin{split}
        j_2 \leq
          &\frac{c}{t} \int^\delta_0 \frac{d\xi_1}{\xi_1}
        \left(\frac{\pi}{2} -\arctg
        \frac{\xi^2_1+2\delta \tau +\delta^2}{2\xi_1t} +
        \arctg \frac{\tau^2-\xi_1^2}{2\xi_1t} - \arctg
        \frac{2\tau\delta - \delta^2-\xi^2_1}{2\xi_1t} \right) \\
        \leq
          &\frac{c}{t} \int^\delta_0 \frac{d\xi_1}{\xi_1}
        \left(\frac{\pi}{2} - \arctg \frac{\delta}{\xi_1} + \arctg
        \frac{t}{\xi_1} - \arctg \frac{\delta}{2\xi_1} \right)
        := \frac{c}{t} j_3,
        \end{split}
        \end{equation}
where we have used the monotonicity of $\arctg x$, for example,
$$\arctg \frac{\xi^2_1+2\delta \tau+\delta^2}{2\xi_1 t}
  \geq \arctg\frac{\delta}{\xi_1},$$
etc., and the relation $\frac{\tau}{t} \to 1$
as $t \to +\infty$, $\tau > t$.

By (6.28),
\begin{equation}
  \frac{\pi}{2} - \arctg \frac{\delta}{\xi_1} \leq \frac{\xi_1}{\delta},
  \ \ \arctg \frac{t}{\xi_1} \leq \frac{\pi}{2} - \frac{\xi_1}{2t},
  \ \ \arctg \frac{\delta}{2\xi_1} \geq \frac{\pi}{2} - \frac{2\xi_1}{\delta}.
  \end{equation}
From (6.35) and (6.36) one gets:
\begin{equation}
        j_2 \leq \frac{c}{t} \int^\delta_0 \frac{d\xi_1}{\xi_1}
        \left(\frac{\xi_1}{\delta} - \frac{\xi_1}{2t} + \frac{2\xi_1}{t}
        \right) =
        \frac{c}{t} \left(1-\frac{\delta}{2t} +\frac{2\delta}{t} \right)
        = \frac{c}{t} \left(1+\frac{3\delta}{t}\right).
        \end{equation}
Thus
\begin{equation}
        j_2 \leq \frac{c}{t}, \quad t \to +\infty.
        \end{equation}

From (6.20), (6.23), (6.25), (6.33) and (6.37) one gets:
\begin{equation}
        |u(x)| \leq c \Vert f\Vert_{L^2(D)}
        \left[\delta + \left(\frac{|\ln \frac{\delta}{2t}|}{t}\right)^
        {\frac{1}{2}} + \frac{1}{t^{\frac{1}{2}}} \right].
        \end{equation}

Choose $\delta = \frac{1}{t}$. Then (6.39) yields
\begin{equation}
        |u(x)| \leq c \Vert f \Vert_{L^2(D)}
        \left(\frac{\ln|\theta|}{|\theta|} \right)^{\frac{1}{2}}, \quad
        |\theta| \to \infty,
        \quad \theta \in M.
        \end{equation}

Estimate (6.11) is proved. \qed
%\end{proof}

{\it Let us prove (6.12).}

Let $L(\xi) = \xi^2 + 2\theta \cdot \xi$, $\partial = -i\nabla$.

Define ${\cal L}(\xi):=\left(\sum_{|j|\geq 0}\left|L^{(j)} (\xi) \right|^2
   \right)^{\frac{1}{2}}$.
Then
$${\cal L}(\xi) = (|\xi^2 +2\theta \cdot \xi|^2 +
4|\xi +\theta|^2 + 36) ^{\frac{1}{2}} \geq |Im \theta| + 3.$$
In \cite{H}, vol. 2, p. 31, it is proved that
$\Vert L^{-1} f \Vert_{L^2(D_1)}
  \leq \frac{1}{\min_{\xi} |{\mathcal L}(\xi)|} \Vert f \Vert_{L^2(D)}$.
Therefore
\begin{equation}
        \begin{split}
        \Vert L^{-1}f \Vert_{L^2(D_1)} \leq
        &\frac{c}{\min_{\xi \in \R^3}|{\cal L}(\xi)|}
        \Vert f \Vert_{L^2(D)} \\
        \leq
        &\frac{c}{|\theta|}
        \Vert f \Vert_{L^2(D)},
        D \subset D_1,
        \quad \theta \in M, \quad |\theta| \to \infty,
        \end{split}
        \end{equation}
where $c = c(D_1, D)>0$ is a constant and we have used the relation
$$c_1|\theta| \leq |Im \theta| \leq |\theta|,
  \quad c_1 >0, \hbox{\ if\ } \theta \in M,
  \quad |\theta | \to \infty.$$

Estimate (6.41) is identical to (6.12). \qed

%%%%%%%%%%%%%%%%%%%%%%%%%%%%%%%%%%%%%%
\subsection{Proof of (2.17).}  
If $\rho$ is defined by (2.4), where $u(x, \alpha)$ solves (1.1) then
$\rho$ solves the equation
\begin{equation}
        \nabla^2 \rho + 2i \theta \cdot \nabla \rho - q(x) \rho=
        q(x) \hbox{\ in\ } \R^3,\quad \theta \in M.
        \end{equation}
Let $h=|\theta|^{-1}$,  $h \to 0$,  $\rho(\xi) := \xi^2+2\beta \cdot \xi$,
$\beta = h \theta$, $\beta \cdot \beta = h^2$, $|\beta| = 1$,
$$N :=\{\xi : \rho(\xi) = 0, \quad \xi \in \R^3 \},
N_h := \{\xi : dist(\xi, N) \leq h, \xi \in \R^3\},
N^\prime_h := \R^3 \backslash N_h,$$
$P = P_1 + iP_2$, $P_1 = Re P$.
Note that $dP_1 \neq 0$ on $N$, where $dP_1$ is the differential of
$P_1$.

Define
\begin{equation}
        F_h u:= \widehat{u} = \frac{1}{(2\pi)^{\frac{3}{2}}} \int_{\R^3} u(x)
        e^{-i\xi \cdot xh^{-1}} dx.
        \end{equation}
Then
\begin{equation}
        F_h(-i\partial_j u(x)) = \xi_j \widehat{u}(\xi) ; \quad
        ih \partial_{\xi_j}
        \widehat{u} (\xi) = \widehat{x_ju}.
        \end{equation}
Denote
\begin{equation}
        \Vert \rho \Vert_a := \Vert \rho \Vert_{L^2(B_a)}, \quad
        \Vert \rho \Vert := \Vert \rho \Vert_{L^2(\R^3)}, \quad
        \Vert \rho \Vert_{a,b} =
        \Vert \rho \Vert_{L^2(B_a \backslash B_a)}, \quad
        b >a,
        \end{equation}
\begin{equation}
        \Vert g (<hD>) \rho \Vert :=
        \Vert g (\sqrt{1 + \xi^2}) \widehat{\rho} (\xi) \Vert, \quad
        D = -i\nabla.
        \end{equation}
The following Hardy-type inequality will be useful:

If $f(t) \in C^1(-h, h)$, $f(0)=0$, then
\begin{equation}
        \int^h_{-h} t^{-2} |f(t)|^2 dt \leq 4 \int^h_{-h} |f^\prime (t)|^2
        dt, \quad h>0.
        \end{equation}

{\it Let us sketch the basic steps of the proof of (2.17)}

\underline{Step 1.} If $\rho \in C^2_0(B_r)$ and
\begin{equation}
        P(hD)\rho := (hD)^2 \rho + 2\beta \cdot hD\rho = -h^2v, \quad
        v \in L^2_0(B_r),
        \end{equation}
where $L^2_0(B_r)$ is the set of $L^2(B_r)$ functions with compact support
in the ball $B_r$, then
\begin{equation}
        h\Vert <hD>^2 \rho \Vert \leq c
        \Vert P(hD)\rho \Vert \quad
        \forall h \in (0, h_0),
        \end{equation}
where $h_0 >0$ is a sufficiently small number.

\underline{Step 2.} Let $A_1$ be a bounded domain with a smooth
boundary and
$A \subset A_1$, $\eta \in C^\infty_0(A_1)$, $0 \leq \eta \leq 1$,
$\eta(x) =1$ in $A$, $A$ is a strictly inner subdomain of $A_1$.

If
\begin{equation}
        P(hD) \rho = 0 \quad \hbox{\ in\ } \quad A_1,
        \end{equation}
then
\begin{equation}
        h \Vert D\rho \Vert_{L^2(A)} \leq c \Vert \rho \Vert_{L^2(A_1)}.
        \end{equation}

\underline{Step 3.} Write (6.42) as
\begin{equation}
        P(hD)\rho = -h^2(q\rho +q).
        \end{equation}
Let
$$\eta \in C^\infty_0 (B_b), \quad 0 \leq \eta (x) \leq 1,$$
 $$\eta (x) = 1 \quad x\in B_{a_1}, \quad a < a_1 < b.$$ 
Then
\begin{equation}
        P(\eta \rho) = (P\eta - \eta P) \rho - h^2\eta (q\rho + q), \quad
        P = P(hD).
        \end{equation}
Apply (6.49) to (6.53) and get
\begin{equation}
        \begin{split}
        h
        &\Vert <hD>^2(\rho \eta)\Vert \leq c
        \Vert (P\eta - \eta P) \rho \Vert + \\
        ch^2
        &\Vert q \Vert_{L^\infty(B_a)}
        \Vert \rho \Vert_{L^2(B_a)}
        + ch^2
        \Vert q \Vert_{L^2(B_a)}.
        \end{split}
        \end{equation}
Since $\eta = 1$ in $B_a$, one gets:
\begin{equation}
        h\Vert \rho \Vert_a \leq h \Vert <hD>^2 (\eta \rho)
        \Vert \leq c h^2 \Vert \rho \Vert_a + ch^2 + c \Vert
        (P\eta - \eta P)\rho \Vert.
        \end{equation}
So
\begin{equation}
        \Vert \rho \Vert_a \leq ch + ch^{-1}
        \Vert(P\eta - \eta P) \rho \Vert.
        \end{equation}

Since $D\eta = 0$ in $B_a$ one gets:
\begin{equation}
        \begin{split}
        &\Vert \left(P \eta - \eta P\right) \rho \Vert =
        \Vert \rho (hD)^2 \eta + 2h^2 D\eta \cdot D\rho + 2h\rho \beta 
        \cdot D\eta \Vert \\
        &\leq c(h^2+h) \Vert \rho \Vert_{a_1,b}
        + ch^2 \Vert D \rho \Vert_{a_1, b}.
        \end{split}
        \end{equation}

Using (6.57), one gets
\begin{equation}
        h \Vert D \rho \Vert_{a_1, b} \leq c
        \Vert \rho \Vert_{a_1 -\varepsilon, b+\varepsilon}.
        \end{equation}

From (6.58), (6.57) and (6.55) one obtains:
\begin{equation}
        \Vert \rho \Vert_a \leq c
        \left(h +\Vert \rho \Vert_{a_1-\varepsilon, b +\varepsilon}
        \right).
        \end{equation}
Since $\varepsilon >0$ is arbitrarily small, the desired inequality (2.17)
follows. \qed

{\it To complete the proof one has to prove} (6.49) and (6.51).

%%%%%%%%%%%%%%%%%%%%%%%%%%%%%%%%%%%%%%
\subsection{Proof of (6.49).}

Write (6.49), using Parseval's equality, as
\begin{equation}
        h \Vert (1+|\xi|^2) \widehat{\rho} \Vert \leq c
        \Vert P(\xi) \widehat{\rho} \Vert.
        \end{equation}
If $\xi \in N^\prime_h$, then $h(1+|\xi|^2) \leq c |P(\xi)|$, so
\begin{equation}
        \begin{split}
        &h^2 \int_{N^\prime_h} (1 +|\xi|^2)^2 |\widehat{\rho} (\xi)|^2
        d\xi \leq c^2 \int_{N^\prime_h} |P(\xi)|^2
        |\widehat{\rho}(\xi)|^2 d\xi   \\
        &\leq c^2
        \int_{\R^3} |P(\xi) \widehat{\rho} (\xi)|^2 d\xi =
        c^2 \int_{\R^3} |P(h D)\rho |^2dx.
        \end{split}
        \end{equation}
If $\xi \in N_h$, then use the local coordinates in which the set $N$ is
defined by the equations:
\begin{equation}
        t =0, \quad \xi_1 =0, \quad t = P_1(\xi),
        \end{equation}
and the $\xi_1$-axis is along vector $\mu$ defined by the equation
$\beta = m +i\mu$. Since $dP_1 \neq 0$ on $N$, these local coordinates can
be defined.

Put $f:= P_1(\xi) \widehat{\rho}(\xi)$. Then $f=0$ at $t=0$,
$f \in C^\infty (\R^3)$ if $\rho(x)$ has compact support, and (6.47) yields:
\begin{equation}
        \int^h_{-h} |\widehat{\rho} (\xi)|^2 dt \leq 4 \int^h_{-h}
        |f^\prime_t|^2 dt.
        \end{equation}
Integrating (6.63) over the remaining variables, one gets:
\begin{equation}
        \int_{N_h} |\widehat{\rho}(\xi)|^2 d\xi \leq c \int_{N_h}
        \left|\nabla_\xi \left(P_1(\xi) \widehat{\rho} (\xi)\right)
        \right|^2 d\xi \leq
        c \int_{\R^3} \left|\nabla_\xi
        \left(P_1(\xi) \widehat{\rho}(\xi)\right)
        \right|^2 d\xi.
        \end{equation}
Since $N_h$ is compact, one has
\begin{equation}
        h^2 \int_{N_h} (1+|\xi|^2)^2 |\widehat{\rho}(\xi)|^2 d\xi \leq ch^2
        \int_{N_h} |\widehat{\rho}(\xi)|^2 d\xi.
        \end{equation}
Using Parseval's equality, S. Bernstein's inequality for the derivative of
entire functions of exponential type, and the condition
$\supp \rho(x) \subset B_r$, one gets:
\begin{equation}
        \begin{split}
        h^2 \int_{\R^3} \left| \nabla_\xi \left(P_1(\xi) \widehat{\rho}(\xi)
        \right) \right|^2 d\xi =\int_{\R^3} |x|^2 |P_1 (hD) \rho (x)|^2 dx
        = r^2 \int_{B_r} \left|P_1(hD) \rho(x)
        \right|^2 dx \\
        \leq r^2 \int_{\R^3} \left|P(hD) \rho(x)\right|^2 dx.
        \end{split}
        \end{equation}
From (6.64)-(6.66) it follows that
\begin{equation}
        h^2 \int_{N_h} (1+|\xi|^2)^2 |\widehat{\rho}(\xi)|^2 d\xi \leq c
        \int_{\R^3} |P(hD) \rho(x)|^2dx.
        \end{equation}
Inequality (6.49) is proved. \qed

%%%%%%%%%%%%%%%%%%%%%%%%%%%%%%%%%%%%%%
\subsection{Proof of (6.51).}
Multiply (6.50) by $\eta \overline{\rho}$, take the real part and integrate
by parts to get:
\begin{equation}
        \begin{split}
        h \int_{A_1} \eta |\nabla \rho|^2 dx
         &= -\frac{h}{2} \int_{A_1}
        \left(\overline{\rho} \nabla \rho + \rho \nabla \overline{\rho}
        \right) \nabla \eta dx + 2 Re  \left(i\beta_j \int_{A_1} \rho_j
        \overline{\rho} \eta dx \right)\\
         &=
        \frac{h}{2} \int_{A_1} |\rho|^2 \nabla^2 \eta dx + 2 Re
        \left(i\beta_j \int_{A_1} \rho_j \overline{\rho} \eta dx \right),
        \end{split}
        \end{equation}
where
$\rho_j := \frac{\partial \rho}{\partial x_j}$ and summation is done
over the repeated indices.

One has
\begin{equation}
        |\nabla^2 \eta |\leq c, \quad
        |\beta_j| \leq 1, \quad
        |2\rho_j \overline{\rho}|
        \leq \frac{h}{2} |\rho_j|^2 + \frac{2}{h}
        |\rho|^2.
        \end{equation}
From (6.69) and (6.68) one gets:
\begin{equation}
        h \int_{A_1} \eta |\nabla \rho|^2 dx \leq ch \int_{A_1}
        |\rho|^2dx +  \frac{h}{2} \int_{A_1} \eta |\nabla \rho|^2 dx +
        \frac{2}{h}
        \int_{A_1} \eta |\rho|^2dx.
        \end{equation}
Thus
\begin{equation}
        h^2 \int_A |\nabla \rho |^2 dx \leq
        h^2 \int_{A_1} \eta |\nabla \rho|^2 dx \leq c \int_{A_1} |\rho|^2 dx.
        \end{equation}
Inequality (6.51) is proved. \qed

\vspace{.1in}
{\it Let us prove that}
\begin{equation}
        \Vert \psi (x, \theta)- \int_{S^2} u(x, \alpha) \nu_\varepsilon
        (\alpha) d \alpha \Vert_{L^2(D)} \leq \varepsilon, \quad
        \theta \in M, \quad
        |\theta| \to \infty,
        \end{equation}
{\it implies}
\begin{equation}
        \Vert \nu_\varepsilon \Vert_{L^2(S^2)} \geq c
        e^{\frac{\kappa d}{2}}, \quad
        \kappa = |Im \theta|, \quad
        d = diam D, \quad
        |\theta| \to +\infty.
        \end{equation}
Indeed, (6.72), (1.19) and (1.20) imply:
\begin{equation}
        \Vert \int_{S^2} u(x, \alpha) \nu_\varepsilon (\alpha)
        d \alpha \Vert_{L^2(D)} \geq \Vert \psi(x, \theta)
        \Vert_{L^2(D)} -\varepsilon \geq ce^{\frac{\kappa d}{2}}, \quad
        \theta \in M, \quad |\theta| \gg  1.
        \end{equation}
If (6.73) is false for some $\varepsilon >0$, then there is a sequence
$\theta_n \in M$, $|\theta_n| \to \infty$, such that
\begin{equation}
        \Vert \nu_\varepsilon \Vert_{L^2(S^2)}
        e^{-\frac{\kappa_nd}{2}} \to 0, \quad
        n \to \infty.
        \end{equation}
This contradicts (6.74) since (6.75) implies
\begin{equation}
        \Vert \int_{S^2} u(x, \alpha) \nu_\varepsilon (\alpha) d \alpha
        \Vert_{L^2(D)} \leq c \Vert \nu_\varepsilon (\alpha)
        \Vert_{L^2(S^2)} = o\left(e^{\frac{\kappa_nd}{2}} \right)
        \hbox{\ as\ } n \to \infty.
        \end{equation}
Therefore estimate (6.73) is proved. \qed

%%%%%%%%%%%%%%%%%%%%%%%%%%%%%%%%%%%%%%
\subsection{Proof of (2.13).}
One has
$$\int_{S^2} u(x, \alpha) \nu(\alpha) d\alpha = e^{i\theta \cdot x}
(1+\rho),$$
where
$$\rho := e^{-i\theta \cdot x} \int_{S^2} u(x, \alpha)\nu(\alpha)
d\alpha -1,$$
$$\psi(x, \theta) = e^{i\theta \cdot x} (1+R),\quad
\Vert R \Vert_{L^2(B_{b_1})} \leq \frac{c}{|\theta|},\quad
\theta \in M, |\theta| \gg 1,$$
where $b_1 >b$.

By (1.18), there exist a $\nu(\alpha)$ such that
$$\Vert e^{i\theta \cdot x} (1+\rho) - e^{i\theta \cdot x} (1+R)
\Vert_{L^2(B_{b_1})} \leq \frac{e^{-\kappa b_1}}{\kappa}, \quad
\kappa = |Im \theta|.$$

Therefore
$$\Vert (\rho -R) e^{i\theta\cdot x} \Vert_{L^2(B_{b_1})} \leq
\frac{e^{-\kappa b_1}}{\kappa},$$
so that
$$e^{-\kappa b_1} \Vert \rho - R \Vert_{L^(B_{b_1})}
 \leq  \frac{e^{-\kappa b_1}}{\kappa},$$
and
$$\Vert \rho -R \Vert_{L^2(B_{b_1})} \leq \frac{1}{\kappa}.$$

This implies
$$\Vert \rho \Vert_{L^2(B_{b_1})} \leq \Vert \rho -R
\Vert_{L^2(B_{b_1})}+\Vert R \Vert_{L^2(B_{b_1})} \leq
\frac{c}{|\theta|}.$$

Thus, inequality (2.13) follows. We claim that
%in fact the infimum in (2.12) 
$\Vert \rho \Vert_{L^2(B_b)}$
is of order $O(\frac 1{|\theta|})$.

Using the above inequalities, one gets:
$$e^{-b\kappa} \Vert \rho -R \Vert_{L^2(B_b)} \leq
\Vert (\rho -R) e^{i\theta \cdot x} \Vert_{L^2(B_b)} \leq
\Vert(\rho -R)e^{i\theta \cdot x} \Vert_{L^2(B_{b_1})} \leq
\frac{e^{-\kappa b_1}}{\kappa}.$$

Thus
$$\Vert \rho -R\Vert_{L^2(B_b)} \leq \frac{e^{-(b_1-b)\kappa}}
{\kappa}.$$
Recall that $c_1 |\theta| \leq \kappa \leq |\theta|$, $0< c_1< \frac{1}{2}$,
as $|\theta| \to \infty$, $\theta \in M$. Therefore,
$$\Vert \rho \Vert_{L^2(B_b)} \geq
\Vert R \Vert_{L^2(B_b)} - \Vert \rho -R \Vert_{L^2(B_b)}
\geq \frac{c}{|\theta|} - \frac{e^{-\gamma \kappa}}{\kappa}, \quad
\gamma = b_1-b >0.$$

Thus, the above claim
is verified, since, as $|\theta| \to \infty$, $\theta
\in M$,
one has $\frac{|\theta|}{\kappa} \to \sqrt{2}$ and
$\frac{e^{-\gamma \kappa}}{\kappa} = o\left(\frac{1}{|\theta|}\right)$. \qed

{\it Uniqueness class for the solution to the equation $L\rho =0$.}
\begin{equation}
        L\rho := (\nabla^2 + 2i\theta \cdot \nabla) \rho = 0
        \hbox{\ in\ } \R^3, \quad
        \int_{\R^3} |\rho(x)|^2 (1+|x|^2)^\l dx < \infty, \quad
        -1 < \l < 0.
        \end{equation}
Taking the distributional Fourier transform of (6.77) one gets:
\begin{equation}
        L(\xi) \widetilde{\rho} = (\xi^2+2\theta \cdot \xi)
        \widetilde{\rho} = 0.
        \end{equation}
Thus $\supp \widetilde{\rho} = C_\tau := \{\xi : \xi \in \R^3, \quad
L(\xi) = 0\}$, and $C_\tau$ is the circle (6.19). By theorem 7.1.27 in
[\cite{H}, vol 1, p.174] one has:
\begin{equation}
        \int_{C_\tau} |\widetilde{\rho}|^2 ds \leq c
        \lim_{r \to \infty} \sup \left(\frac{1}{R^2} \int_{|x| \leq R}
        |\rho(x)|^2 dx \right).
        \end{equation}
Using (6.77) we derive for $-1 < \l < 0$:
\begin{equation}
        \begin{split}
        \infty > c >
        &\int_{\R^3} |u|^2(1+|x|^2)^\l dx \geq \int_{|x| \leq R}
        \frac{|u|^2dx}{(1+|x|^2)^{|\l|}} \\
        \geq
        &\frac{1}{(1+R^2)^{|\l|}}
        \int_{|x| \leq R}
        |u|^2dx \geq \frac{c}{R^{2|\l|}}
        \int_{|x| \leq R} |u|^2dx.
        \end{split}
        \end{equation}
Combining (6.79) and (6.80) one gets
$$ \int_{C_\tau} |\widetilde{\rho}|^2 ds \leq c \lim_{R \to \infty}
\sup \frac{R^{2|\l|}}{R^2} = 0, \quad |\l| < 1.$$
Thus $\widetilde{\rho}(\xi) = 0$, as claimed. \qed

The above argument is valid in $\R^n$, $n \geq 2$. It was used in
\cite{SU} and  \cite {R2}.

%%%%%%%%%%%%%%%%%%%%%%%%%%%%%%%%%%%%%%
\subsection{Proof of (2.23)}
Let $\Vert \nu_\varepsilon \Vert :=
\Vert \nu_\varepsilon \Vert_{L^2(S^2)}$
and $m(\varepsilon, \theta) := \inf \Vert \nu \Vert$
where the infimum is taken over all $\nu \in L^2(S^2)$ such that (6.72) holds.

We wish to prove that
\begin{equation}
        m(\varepsilon, \theta) \leq ce^{c|\theta|\ln |\theta|}
        \hbox{\ as\ } |\theta | \to \infty, \quad \theta \in M, \quad
        \varepsilon = \frac{e^{-b\kappa}}{\kappa}, \quad
        b > a,
        \end{equation}
where $\theta \in M$, $|\theta| \to \infty$,
$\kappa = |Im \theta|$, and $c >0$ stands for various
constants.

Let us describe the steps of the proof.

\underline{Step 1.}
Prove the estimate
\begin{equation}
        m(\varepsilon, \theta) \leq ce^{\kappa r}
        \left(\frac{2n(\varepsilon)}{er}\right)^{n(\varepsilon)}
        n^2(\varepsilon), \quad r \geq b, \quad
        \theta \in M, \quad \varepsilon > 0,
        \end{equation}
where
\begin{equation}
        \ln (n(\varepsilon)) = \ln(|\ln \varepsilon|) [1 + o(1)], \quad
        \varepsilon \to +0.
        \end{equation}
The choice of $n(\varepsilon)$ in (6.83) is justified below (see (6.97))
and 
estimate (6.82) is proved also below.

\underline{Step 2.}
Minimize the right-hand side of (6.82) with respect to $r \geq b$ to get
\begin{equation}
        m(\varepsilon, \theta) \leq c(2\kappa)^{n(\varepsilon)}
        n^2(\varepsilon).
        \end{equation}

The minimizer is $r =\frac{n(\varepsilon)}{\kappa}$.

\underline{Step 3.}
Take $\varepsilon = \varepsilon (\theta) = \frac{e^{-\kappa}}{\kappa}$,
$\kappa \to \infty$, in (6.84). Then
\begin{equation}
        \ln n = \ln (\kappa b + \ln \kappa) [1+ o(1)] = (\ln \kappa)
        \left[1 +O\left(\frac{1}{\ln \kappa} \right) \right],
        \quad \kappa \to +\infty,
        \end{equation}
so, for $\varepsilon = \frac{e^{-\kappa b}}{\kappa}$ one has:
\begin{equation}
        c_1 \kappa \leq n \leq c_2 \kappa, \quad
        \kappa \to +\infty, \quad
        c_1 > 0.
        \end{equation}

From (6.84) and (6.85) one gets:
\begin{equation}
        m(\theta) = m(\varepsilon (\theta), \theta) \leq c
        e^{c|\theta| \ln |\theta|}, \quad
        |\theta| \to \infty, \quad
        \theta \in M.
        \end{equation}

Estimate (6.81) is obtained.

{\it Proof of (6.82).}
Since $u(x, \alpha) = (I +T_1) e^{i \alpha \cdot x}$ where $I + T_1$ is a
bijection of $C(B_b)$ onto $C(B_b)$, inequality (6.72) with $D = B_b$
is equivalent to
\begin{equation}
        \Vert (I+T_1)^{-1} \psi -\int_{S^2} e^{i \alpha \cdot x}
        \nu_\varepsilon (\alpha) d \alpha \Vert_{L^2(B_b)}
        \leq c \varepsilon,
        \end{equation}
where $c = \const >0$ does not depend on $\varepsilon$ and $\theta$,
$(I +T_1)^{-1} \psi = (I +T) \psi$,
$$T \psi = \int_{B_a} \frac{e^{i|x-y|}}{4\pi |x-y|} q(y) \psi (y) dy.$$
We take $b >a$, therefore the function $\varphi := \psi + T\psi$, has the
maximal values, as $|\theta| \to \infty$, of the same order of magnitude
as the function $\psi$. The function $\varphi$ solves the equation
\begin{equation}
        \left(\nabla^2 + 1\right) \varphi = 0 \hbox{\ in\ }
        \R^3.
        \end{equation}
Indeed, $(\nabla^2 +1) \varphi = (\nabla^2 +1) \psi - q\psi =
q\psi -q\psi =0$,
as claimed.

Therefore on can write:
\begin{equation}
        \varphi := \varphi (x, \theta) = \sum^\infty_{\l =0} 4\pi i^\l
        \varphi_\l Y_\l (\alpha^\prime) j_\l (r),\quad
        r=|x|, \quad
        \alpha^\prime = \frac{x}{|x|},
        \end{equation}
where $Y_{\l}$ are defined in (1.26), $j_\l(r)$ are defined in (1.29),
and $\varphi_\l = \varphi_\l (\theta)$ are some coefficients.

It is known that
\begin{equation}
        e^{i\alpha \cdot x} = \sum^\infty_{\l = o} 4 \pi i^\l
        \overline{Y_\l(\alpha)} Y_\l (\alpha^\prime) j_\l (r),
        \end{equation}
so
\begin{equation}
        \int_{S^2} e^{i\alpha \cdot x} \nu_\varepsilon (\alpha) d \alpha =
        \sum^\infty_{\l =0} 4 \pi i^\l \nu_{\varepsilon \l}
        Y_\l (\alpha^\prime) j_\l(r),
        \end{equation}
where $\nu_{\varepsilon \l} = (\nu_\varepsilon, Y_\l)_{L^2(S^2)}$.

Choose
\begin{equation}
        \nu_{\varepsilon \l} = \varphi_\l \quad \hbox{\ for\ } \quad \l \leq
        n(\varepsilon), \quad \nu_{\varepsilon \l}=0 \quad \hbox{\ for\ }
        \quad \l > n(\varepsilon),
        \end{equation}
where $n(\varepsilon)$ is the same as in (6.83).

Then (6.88) implies:
\begin{equation}
   \begin{split}
    & \Vert \varphi - \int_{S^2} e^{i \alpha \cdot x}
        \nu_\varepsilon (\alpha) d\alpha \Vert^2_{L^2(B_b)}\\
    & \qquad=
        \sum^\infty_{\l =n(\varepsilon) +1} 16\pi^2 
        \int^b_0 r^2 |j_\l(r)|^2 dr |\varphi_\l|^2 
    \leq c \sum^\infty_{\l = n(\varepsilon)+1} \frac{1}{\l^2}
        |\varphi_\l|^2
        \left(\frac{eb}{2\l}\right)^{2\l} < c\varepsilon,
        \end{split}
        \end{equation}
where formula (1.29) was used.

From (6.92) and formula (6.91) with $\alpha = \theta \in M$, one gets:
\begin{equation}
        \Vert \nu_\varepsilon \Vert^2 = \sum^{n(\varepsilon)}_{\l =0}
        |\varphi_\l|^2 \leq c \sum^{n(\varepsilon)}_{\l =0}
        \sum^\l_{m= -\l} \left| Y_\l(\theta)\right|^2 \leq
        cn^2(\varepsilon)
        \frac{e^{2\kappa r}}{|j_{n(\varepsilon)}(r)|^2}, \quad
        \forall r >0,
        \end{equation}
where we have used the formula $\sum^n_{\l=0}$
$\sum^\l_{m = -\l} = (n+1)^2$, we estimated $|\varphi_\l|$ by the
coefficient
$|(e^{i\theta \cdot x}, Y_\l)_{L^2(S^2)}|^2 = 16\pi^2|Y_\l(\theta)|^2$
of the main term of $\varphi$, that is, the function $e^{i\theta \cdot x}$,
we used estimate (1.28), which gives
$|\varphi_\l|^2 \leq c\frac{e^{2\kappa r}}{|j_\l(r)|^2}$ $\forall r > 0$,
and we replaced $|j_\l (r)|$ by $|j_{n(\varepsilon)} (r)|$, the smaller
quantity.

Choose $r>b$ and use (1.29) to get the inequality:
\begin{equation}
        \sum^\infty_{\l = n(\varepsilon)+1} |\varphi_\l|^2
        \left(\frac{eb}{2\l}\right)^{2\l} \leq
        \sum^\infty_{\l = n(\varepsilon)+1} e^{2\kappa r}
        \left(\frac{b}{r}\right)^{2\l} \leq c_1e^{2\kappa r}
        \left(\frac{b}{r}\right)^{2n(\varepsilon)} < c\varepsilon, \quad
        r >b,
        \end{equation}
which implies (6.94). Thus (6.94) holds if
$e^{\kappa r} \left(\frac{b}{r}\right)^{n(\varepsilon)}
\leq c\sqrt{\varepsilon}$, where $c$ stands for various
constants. One has
$\min_{r >b} e^{\kappa r} \left(\frac{b}{r}\right)^n = e^n
\left(\frac{b\kappa}{n}\right)^n$,
and the minimizer is $r = \frac{n}{\kappa}$. Consider therefore the
equation $e^n \left(\frac{b \kappa}{n}\right)^n = c \sqrt{\varepsilon}$
and solve it asymptotically for $n= n(\varepsilon)$ as
$\varepsilon \to 0$, where $\kappa >1$ is arbitrary large but fixed.
Taking logarithm, one gets
$\ln c -\frac{1}{2} \ln \frac{1}{\varepsilon} = n-n\ln n+ n\ln (b\kappa)$.

Thus $|\ln \varepsilon|=\ln \frac{1}{\varepsilon} = 2n\ln n[1+o(1)]$,
and
\begin{equation}
        \ln |\ln \varepsilon| = (\ln n) (1 +o(1)), \quad
        \varepsilon \to +0.
        \end{equation}

Hence, we have justified (6.83).

From (6.94), (6.96) and (1.29), one gets
\begin{equation}
        \Vert \nu_\varepsilon \Vert \leq c \quad n^2(\varepsilon)
        \frac{e^{\kappa r}(2n(\varepsilon))^{n(\varepsilon)}}
        {(er)^{n(\varepsilon)}} \quad
        \forall r > b, \quad
        \kappa = |Im \theta|, \quad \theta \in M.
        \end{equation}
Estimate (6.82) is established. \qed

%%%%%%%%%%%%%%%%%%%%%%%%%%%%%%%%%%%%%%
\subsection{Proof of (1.30).}
Let $G_j$ be the Green function corresponding to $q_j(x)$, $j=1,2$. By
Green's formula one gets
\begin{equation}
        G_2 (x,y) - G_1(x,y) = \int_{B_a} p(z) G_1(x,z) G_2(z, y) dz, \quad
        p := q_1(x) - q_2(x).
        \end{equation}
Take $|y| \to \infty$, $\frac{y}{|y|} = - \alpha$ and use (1.7) to get:
\begin{equation}
        u_2(x, \alpha) - u_1(x, \alpha) = \int_{B_a} p(z)
        G_1 (x,z) u_2(z, \alpha) dz.
        \end{equation}
Take $|x| \to \infty$, $\frac{x}{|x|} = \alpha^\prime$, use (1.7) and (1.2)
and get:
\begin{equation}
        A_2(\alpha^\prime, \alpha) - A_1(\alpha^\prime, \alpha) =
        \frac{1}{4\pi} \int_{B_a} u_1(z, -\alpha^\prime) u_2(z, \alpha)
        p(z) dz.
        \end{equation}
Since $A(\alpha^\prime, \alpha) = A(-\alpha, -\alpha^\prime)$,
formula (6.101) is equivalent to (1.30). \qed

%%%%%%%%%%%%%%%%%%%%%%%%%%%%%%%%%%%%%%%%%
\section{Construction of the Dirichlet-to-Neumann map from the scattering
data and vice versa.}%7
Consider a ball $B_a \supset D = \supp q(x)$ and assume that the problem
\begin{equation}
        \left[\nabla^2 +1 - q(x)\right] w = 0 \hbox{\ in\ }
        B_a, \quad w = f \hbox{\ on\ }
        S_a := \partial B_a,
        \end{equation}
is uniquely sovable for any $f \in H^{\frac{3}{2}}(S_a)$, where
$H^\l (S_a)$ is the Sobolev space.

Then the $D-N$ map is defined as
\begin{equation}
        \Lambda : f \to w_N
        \end{equation}
where $w_N$ is the normal derivative of $w$ on $S_a$, $N$ is the
normal to $S_a$ pointing into $B_a^\prime := \R^3 \backslash B_a$.

If $\Lambda$ is known, then $q(x)$ can be found as follows.

The special solution (1.19)-(1.22) satisfies the equation:
\begin{equation}
        \psi (x) = e^{i\theta \cdot x} - \int_{B_a} G(x-y) q(y) \psi(y) dy,
        \end{equation}
where $G(x) = e^{i\theta \cdot x} G_0(x)$ and
$\nabla^2 G(x) + G(x) = -\delta(x)$ in $\R^3$. Thus

\begin{equation}
        \nabla^2 G_0 +2i \theta \cdot \nabla G_0 = -\delta (x),
        \end{equation}
so that $G_0(x-y)$ is the Green function of the operator $L$, see (6.10),
that is
\begin{equation}
        G_0(x) = \frac{1}{(2\pi)^3} \int_{\R^3}
        \frac{e^{i\xi \cdot x}d\xi}{\xi^2 +2\xi \cdot \theta}.
        \end{equation}

The function $G(x)$ can be considered known.

Since $q\psi = (\nabla^2 +1) \psi$, one can write, for $x \in B_a^\prime$,
\begin{equation}
        \begin{split}
        &\int_{B_a} G(x-s) q(y) \psi(y) dy \int_{B_a} G(\nabla^2 + 1)\psi dy\\
        &=\int_{S_a}
        \left[G(x-s) \psi_N (s) - G_N(x-s) \psi (s)\right] ds =
        \int_{S_a} G(x-s) (\Lambda - \Lambda_0) \psi (s) ds  \\
        &+\int_{S_a} \left[G(x-s) \Lambda_0\psi - G_N(x-s) \psi\right] ds
        =\int_{S_a} G(x-s) (\Lambda - \Lambda_0) \psi (s) ds.
        \end{split}
        \end{equation}
Here $\Lambda _0$ is $\Lambda$ for $q(x) = 0$, we have used Green's formula
and took into account that
$$ \int_{S_a}\left[G(x-s) \Lambda_0 \psi - G_N(x-s) \psi \right] ds=
\int_{B_a} \left[G(\Delta +1) \varphi - \varphi(\Delta +1) G\right]
dy =0,$$
where $\varphi$ solves problem (7.1) with $q(x) = 0$ and
$\varphi = f$ on
$S_a$.

From (7.3) and (7.6) taking $x \to s \in S_a$ one gets a linear Fredholm-
type equation for $\psi|_{S_a}$:
\begin{equation}
        \psi (s) = e^{i\theta \cdot s} - \int_{S_a} G(s-s^\prime)
        (\Lambda - \Lambda_0) \psi(s^\prime) ds^\prime.
        \end{equation}

If $\Lambda$ is known, one can find from (7.7) $\psi |_{S_a}$ and then find
$q(x)$ using the following calculation.

Define
\begin{equation}
        t(\theta^\prime, \theta) := \int_{B_a} e^{-i\theta^\prime \cdot y}
        q(y) \psi (y, \theta) dy.
        \end{equation}

By Green's formula, as in (7.6), one gets
\begin{equation}
        t(\theta^\prime, \theta) = \int_{S_a} e^{-i\theta^\prime \cdot s}
        (\Lambda - \Lambda_0) \psi (s) ds.
        \end{equation}

From (7.8) one gets, using (1.19), (1.20) and (1.9):
\begin{equation}
        \lim_
        {\begin{array}{ll}
        |\theta| \to \infty \\
        \theta^\prime - \theta = \xi \\
        \theta \in M
        \end{array}}
        t(\theta^\prime, \theta) = \int_{B_a} e^{-i\xi \cdot x}
        q(x)dx := \widetilde{q}(\xi).
        \end{equation}

Therefore the knowledge of $\Lambda$ allows one to recover
$\widetilde{q}(\xi)$ by formula (7.10), but first one has to solve equation
(7.7). We leave to the reader to check that the homogeneous equation (7.7)
has only the trivial solution so that Fredholm-type equation (7.7) is
uniquely solvable in $L^2(S_a)$  (see a proof in \cite{R26}).

Practically, however, there are essential difficulties: a) the function
$G(x,y)$ is not known, analytically and it is difficult to solve equation
(7.7) by this reason, b) the $D-N$ map is not given analytically as well.

{\it Let us show how to construct $\Lambda$ from the scattering amplitude
$A(\alpha^\prime, \alpha)$ and vice versa.} If $\Lambda$ is given then we
have shown how to find $q(x)$ and if $q(x)$ found then
$A(\alpha^\prime, \alpha)$, the scattering amplitude, can be found.

Conversely, suppose $A(\alpha^\prime, \alpha)$ is known. Then the scattering
solution can be calculated in $B_a^\prime$ by formula (1.31).

Let $f \in H^{\frac{3}{2}} (S_a)$ be given, $g(x,y)$ be the Green function
of the operator $-\nabla^2 +q(x) -1$ in $\R^3$ which satisfies the
radiation condition (1.6), and define
\begin{equation}
        w (x) = \int_{S_a} g(x, s) \sigma(s) ds,
        \end{equation}
such that
\begin{equation}
        w = f \hbox{\ on\ } S_a.
        \end{equation}

Since $(\nabla^2 +1) w = 0$ in $B_a^\prime$, $w = f$ on $S_a$
and $w$ satisfies (1.6), one can find $w$ in $B_a^\prime$
explicitly:
\begin{equation}
        w (x) = \sum^\infty_{\l =0} \frac{f_\l}{h_\l(a)}
        Y_\l(\alpha^\prime) h_\l (r), \quad
        r \geq a, \quad
        z = |x|, \quad \alpha^\prime = \frac{x}{r},
        \end{equation}
where $f_\l$ are the Fourier coefficients of $f$:
\begin{equation}
        f(s) = \sum^\infty_{\l =0} f_\l Y_\l (\alpha^\prime), \quad
        s \in S_a.
        \end{equation}
Therefore the function
$$
w^-_N = \lim_{{|x| \to a}, {x \in B_a^\prime}}
\frac{\partial w (x)}{\partial r}
$$ is known.
By the jump formula for single-layer potentials one has (\cite{R7}, p. 14)
\begin{equation}
        w^+_N = w^-_N + \sigma.
        \end{equation}
The map $\Lambda : f \to w^+_N$ is constructed as soon as we find
$\sigma(s)$, because $w^-_N$ is already found.

To find $\sigma$, consider the asymptotics of $w (x)$ as
$|x| \to \infty$, $\frac{x}{|x|} = \beta$. Using (1.7) and (7.11), one gets:
\begin{equation}
        \frac{1}{4\pi} \int_{S_a} u(s, -\beta) \sigma (s) ds = \eta (\beta)
        := \sum^\infty_{\l =0}
        \frac{f_\l Y_\l (\beta)}{h_\l (a)},
        \end{equation}
where we have used (7.13) and the asymptotics
$h_\l (r) \sim \frac{e^{ir}}{r}$ as $r \to +\infty$. As we have already
mentioned, the function $u(s, \alpha^\prime)$ is known explicitly
(see formula (1.31)), and equation (7.16) is uniquely solvable for
$\sigma (s)$. Analytical solution of equation (7.16)
 for $\sigma (s)$ can be obtained as
a series
\begin{equation}
        \sigma (s) = \sum^\infty_{\l=0} \sigma_\l Y_\l (\alpha^\prime),
        \quad \alpha^\prime = \frac{s}{|s|}.
        \end{equation}

Substitute (6.91) with $\alpha = -\beta$ into (1.31), take $r =a$ in (1.31)
and $\alpha^\prime = \frac{s}{a}$, and substitute (1.31) into (7.16).
By our choice of the
spherical harmonics (1.26) both systems
$\{Y_\l\}_{\l = 0,1,2, \dots}$ and
$\{\overline{Y_\l}\}_{\l =0,1,2,\dots}$
form orthonormal bases of $L^2(S^2)$. Therefore one gets:
\begin{equation}
        \begin{split}
        &\frac{1}{4\pi} \sum^\infty_{\l = 0} 4\pi i^\l
        \overline{Y_\l} (-\beta) j_\l(a) a^2 \int_{S^2} Y_\l (\alpha^\prime)
        \sigma (a \alpha^\prime) d\alpha^\prime \\
        & +\frac{1}{4\pi} \sum^\infty_{\l = 0} A_\l(-\beta) h_\l(a) a^2
        \int_{S^2} Y_\l (\alpha^\prime) \sigma (a\alpha^\prime)
        d\alpha^\prime \\
        &=\sum^\infty_{\l =0} \frac{f_\l Y_\l (\beta)}{h_\l (a)}.
        \end{split}
        \end{equation}
Denote
\begin{equation}
        \int_{S^2} \sigma (a \alpha^\prime) Y_{\l, m} (\alpha^\prime)
        d\alpha^\prime := \sigma_{\l m}.
        \end{equation}

Using (1.26) one gets:
\begin{equation}
        \begin{split}
        Y_{\l, m} (-\beta)
          &= (-1)^\l Y_{\l, m} (\beta),\quad
        \overline{Y_{\l, m}(-\beta)} = (-1)^\l
        \overline{Y_{\l, m}(\beta)} \\
          &= (-1)^{\l +\l +m} Y_{\l, -m}(\beta)
        = (-1)^m Y_{\l, -m} (\beta).
        \end{split}
        \end{equation}
Also define $A_{\l m, \l^\prime m^\prime}$ by the formula:
\begin{equation}
        A_{\l,m}(-\beta) = \sum_{\l^\prime, m^\prime}
        A_{\l,m, \l^\prime m^\prime}
        Y_{\l^\prime, -m^\prime} (\beta).
        \end{equation}
The above definition differs from (1.36) and is used for convenience in
this section.
  
Equating the coefficients in front of $Y_{\l, -m}(\beta)$ in (7.18) one gets
\begin{equation}
        i^\l(-1)^m j_\l (a)a^2 \sigma_{\l m} + \frac{a^2}{4\pi}
        \sum^\infty_{\l^\prime =0} \sum^{\l^\prime}_{m^\prime = -\l^\prime}
        A_{\l^\prime m^\prime, \l,m} h_{\l^\prime} (a) 
        \sigma_{\l^\prime m^\prime} =
        \frac{f_{\l,-m}}{h_\l (a)},
        \end{equation}
or
\begin{equation}
        \sigma_{\l m} + \frac{(-1)^m(-i)^\l}{4\pi j_\l(a)}
        \sum^\infty_{\l^\prime = 0} \sum^{\l^\prime}_{m^\prime = -\l^\prime}
        A_{\l^\prime m^\prime, \l,m}
        h_{\l^\prime} (a) \sigma_{\l^\prime m^\prime} =
        \frac{f_{\l, -m}(-1)^m(-i)^\l}{a^2j_\l(a) h_\l(a)}.
        \end{equation}
The matrix of the linear system (7.23) is ill-conditioned (see \cite{R26},
where estimates of the entries of the matrix of (7.19) are obtained and the
case of the noisy data is mentioned). \qed

Finally let us show (see \cite{R27}) that {\it it is impossible to get an
estimate}
\begin{equation}
        \Vert Q f \Vert \leq \varepsilon (|\theta|)
        \Vert f \Vert, \quad
        \theta \in M, \quad
        \Vert f \Vert :=
        \Vert f \Vert_{L^2(D)}, \quad
        \varepsilon (t) \to 0 \hbox{\ as\ } t \to +\infty,
        \end{equation}
if
\begin{equation}
        Qf = \int_D \Gamma (x, y, \theta) f(y) dy,
        \end{equation}
where
\begin{equation}
        L\Gamma := (\nabla^2 + 2i\theta \cdot \nabla) \Gamma =
        -\delta(x-y) \hbox{\ in\ } D, \quad
        \theta \in M,
        \end{equation}
\begin{equation}
        \Gamma = 0 \hbox{\ on\ } S:= \partial D,
        \end{equation}
and we assume that
\begin{equation} 
        \hbox{\ the\ } \hbox{\ problem\ }
        L \rho = 0, \quad
        \rho =0 \quad
        \hbox{\ on\ } \quad S \quad \hbox{\ has\ } \hbox{\ only\ }
        \hbox{\ the\ } \hbox{\ trivial\ } \hbox{\ solution\ }.
        \end{equation}
Indeed, choose a $q(x) \in L^\infty (D)$ such that the problem
\begin{equation}
        \left[\nabla^2 +1 - q(x)\right] w = 0 \hbox{\ in\ } D, \quad
        w = 0 \hbox{\ on\ } S,
        \end{equation}
has a non-trival solution.

Define $\rho = e^{-i\theta \cdot x} w$. Then $\rho \not\equiv 0$, and
\begin{equation}
        L\rho - q\rho = 0 \hbox{\ in\ } D, \quad
        \rho = 0 \hbox{\ on\ } S.
        \end{equation}
Because of our assumption (7.24), one gets:
\begin{equation}
        \rho = \int_D \Gamma (x,y) q(y) \rho (y) dy := T\rho.
        \end{equation}
Were (7.24) true, it would imply for $|\theta| \gg  1$, $\theta \in M$,
that the operator $T: L^2(D) \to L^2(D)$ in (7.31) has small norm, so
$\rho = 0$, contrary to our assumption. \qed

%%%%%%%%%%%%%%%%%%%%%%%%%%%%%%%%%%%%%%%
\section{Examples of nonuniqueness for an inverse problem of geophysics.}
%8
%%%%%%%%%%%%%%%%%%%%%%%%%%%%%%%%%%%%%%
\subsection{Statement of the problem.}  
In this section the result from \cite{R21} is presented.

Let  $D \subset {\mathbb R}_+^n := \{ x : x \in {\mathbb R}^n, x_n \geq 0\}$
be a bounded domain, part  $S$  of the boundary  $\Gamma$  of  $D$  is
on the plane  $x_n = 0$, $f(x,t)$  is a source of the wavefield, $c(x) >
0$  is a velocity profile.  The wavefield, e.g., the acoustic pressure,
solves the problem:
\begin{equation}
        c^{-2} (x) u_{tt} - \Delta u = f(x,t) \quad\hbox{in}\quad D \times [0,
        \infty),\quad f(x,t) \not\equiv 0, %\tag 1
        \end{equation}
\begin{equation}
        u_N = 0 \quad\hbox{on}\quad \Gamma %\tag 2
        \end{equation}
\begin{equation}
         u = u_t = 0 \quad\hbox{at}\quad t = 0.  %\tag 3
         \end{equation}
Here  $N$  is the unit outer normal to  $\Gamma$, $u_N$  is the normal
derivative of  $u$  on  $\Gamma$.  If  $c^2 (x)$  is known, then the
direct problem (8.1)-(8.3) is uniquely solvable.  The inverse problem (IP)
we are interested in is the following one:

\vspace{ .2in}

\noindent (IP) {\it Given the data  $u(x,t) \quad \forall x \in S$,
$\forall t > 0$, can one recover  $c^2 (x)$  uniquely?}

\vspace{ .2in}

The basic result is: {\it the answer to the above question
is no}.

An analytical construction is presented of two constant
velocities  $c_j > 0$, $j = 1,2,$ $c_1 \not= c_2$, which can be chosen
arbitrary,
and a source, which is constructed after  $c_j > 0$  are chosen, such
that the solutions to problems (8.1)-(8.3) with  $c^2 (x) = c_j^2$, $j=1,2,$
produce the same surface data on  $S$  for all times:
\begin{equation}
        u_1 (x,t) = u_2 (x,t) \quad \forall x \in S, \quad \forall t > 0.
        %\tag 4
        \end{equation}
The domain  $D$  we use is a box: $D = \{ x : a_j \leq x_j \leq b_j, 1
\leq j \leq n\}$.

This construction is given in the next section. At the end of section 8.2 the
data on  $S$  are suggested, which allow one to uniquely determine  $c^2
(x)$.

\vspace{ .2in}

%%%%%%%%%%%%%%%%%%%%%%%%%%%%%%%%%%%%%%
\subsection{Example of nonuniqueness of the solution to IP.}  

Our construction is valid for any  $n \geq 2$.  For simplicity we take
$n = 2$, $D = \{ x : 0 \leq x_1 \leq \pi, 0 \leq x_2 \leq \pi\}$.  Let
$c^2 (x) = c^2 = \hbox{const} > 0$.  The solution to (8.1)-(8.3) with  $c^2
(x) = c^2 = \hbox{const}$  can be found analytically
\begin{equation}
       u(x,t) = \sum_{m=0}^\infty u_m (t) \phi_m (x), \quad m = (m_1,m_2)
       %\tag 5
       \end{equation}
where
\begin{equation}
        \begin{split}
        \phi_m (x) &= \gamma_{m_1m_2} \cos (m_1 x_1) \cos (m_2 x_2), \\
        \int_D \phi_m^2(x)dx
                   &= 1, \quad \Delta \phi_m +\lambda_m \phi_m = 0,\\
        \phi_{m N}
                  &= 0 \quad\hbox{on}\quad \Gamma,
                  \quad \lambda_m := m_1^2 +m^2_2,     \\
       \gamma_{00} &= \frac 1\pi, \quad \gamma_{m_10} = \gamma_{0m_2} = \frac
       {\sqrt{2}}\pi,
       \end{split}
       \end{equation}
$$\gamma_{m_1m_2} = 2/\pi \ \hbox{if}\quad m_1 > 0 \quad\hbox{and}\quad
m_2 > 0,$$
\begin{equation}
        \begin{split}
        &u_m (t) := u_m (t,c) = \frac c{\sqrt{\lambda_m}} \int_0^t \sin [c
        \sqrt{\lambda_m} (t - \tau)] f_m (\tau) d\tau, \\
        &f_m (t) := \int_D
        f(x,t) \phi_m (x) dx .
        \end{split}
        \tag{$8.6^\prime$}
        \end{equation}
The data are
\begin{equation}
   u(x_1,0,t) = \sum_{m=0}^\infty u_m (t,c) \gamma_{m_1 m_2} \cos (m_1
   x_1). %\tag 7
   \end{equation}
For these data to be the same for  $c = c_1$  and  $c= c_2$, it is
necessary and sufficient that
\begin{equation}
        \sum_{m_2 = 0}^\infty \gamma_{m_1 m_2} u_m (t,c_1) =
        \sum_{m_2 = 0}^\infty \gamma_ {m_1 m_2}u_m (t,c_2), \quad \forall t >
        0,\quad \forall m_1.  %\tag 8
        \end{equation}
Taking Laplace transform of (8) and using (6$^\prime$) one gets an
equation, equivalent to (8.8),
\begin{equation}
        \sum_{m_2=0}^\infty
        \gamma_ {m_1 m_2} \overline{f}_m (p) \left[ \frac {c_1^2}{p^2 + c_1^2
        \lambda_m} - \frac {c_2^2}{p^2 + c_2^2 \lambda_m}\right] = 0, \quad
        \forall p > 0,\quad \forall m_1.  %\tag 9
        \end{equation}

Take  $c_1 \not= c_2$, $c_1, c_2 > 0$, arbitrary and find
$\overline{f}_m (p)$  for which (8.9) holds.  This can be done by
infinitely many ways.  Since (8.9) is equivalent to (8.8), the desired
example of nonuniqueness of the solution to IP is constructed.

Let us give a specific choice: $c_1 = 1$, $c_2 = 2$,
$\overline{f}_{m_1m_2} = 0$  for  $m_1 \not= 0$, $m_2 \not= 1$  or  $m_2
\not= 2$, $\overline{f}_{02} (p) = \frac 1{p + 1}$, $\overline{f}_{01}
(p) = -\frac {p^2 + 1}{(p + 1)(p^2 + 16)}$.  Then (8.9) holds.  Therefore,
if
\begin{equation}
        f(x,t) = \frac {\sqrt{2}}\pi \left[ f_{01} (t) \cos (x_2) + f_{02} (t)
        \cos (2x_2)\right], \quad c_1 = 1, \quad c_2 = 2, %\tag 10
        \end{equation}
then the data  $u_1 (x,t) = u_2 (x,t) \quad \forall x \in S$, $\forall t
> 0$.  In (8.10) the values of the coefficients are
\begin{equation}
        f_{01} (t) = -\frac 2{17} \exp (-t) - \frac {15}{17} \left[ \cos (4t)
        \frac 14 \sin (4t)\right], \quad f_{02} (t) = \exp (-t).  %\tag 11
         \end{equation}

\vspace{ .2in}

\begin{remark}
The above example brings out the question:

What data on
$S$  are sufficient for the unique identifiability of  $c^2 (x)$?

The
answer to this question one can find in \cite{R7} and \cite{R2}.

In particular, if one takes  $f(x,t) = \delta (t) \delta (x - y)$, and
allows  $x$  and  $y$  run through  $S$, then the data  $u(x,y,t) \quad
\forall x,y \in S$, $\forall t > 0$, determine  $c^2 (x)$  uniquely.  In
fact, the low frequency surface data  $\tilde u(x,y,k)$, $\forall x,y
\in S \quad \forall k \in (0,k_0)$, where  $k_0 > 0$  is an arbitrary
small fixed number, determine  $c^2 (x)$  uniquely under mild
assumptions on  $D$  and  $c^2 (x)$.  By  $\tilde u (x,y,k)$  is meant
the Fourier transform of  $u(x,y,t)$  with respect to  $t$.
\end{remark}

\begin{remark}
One can check that the non-uniqueness example with
constant velocities is not possible to construct, as was done above,
if the sources are concentrated on $S$, that is, if $f(x_1,x_2,t)=
\delta (x_2)f_1(x_1,t)$.
\end{remark}

%%%%%%%%%%%%%%%%%%%%%%%%%%%%%%%%%%%%%%%

\section{A uniqueness theorem for inverse boundary
value problem for parabolic equations}%9

Consider the problem:

\begin{equation}
  u_t+Lu=0, \quad x\in D, \quad t\in [0,T],
  \end{equation}
\begin{equation}
 u=0 \quad \hbox {at} \quad t=0
 \end{equation}
\begin{equation}
  u=f(s)\delta (t)\quad \hbox {on} \quad S.
  \end{equation}
Here $\delta(t)$ is the delta-function,
$D$ is a bounded domain in $\R^n, n\geq 3,$
with a smooth boundary $S$, $f\in H^{3/2}(S),$ 
$Lu:=-div [a(x) grad u] +q(x)u$, $a(x)$ and $q(x)$ are
real-valued functions, $q\in L^2(D),$ $0<a_0\leq a(x) \leq a_1$,
where $a_0$ and $a_1$ are positive constants, and
$a(x)\in C^2(\bar D)$, where $\bar D$ is the closure of $D$.
Let $h(s,t):=a(s) u_N$, where $N$ is the unit exterior normal
to $S$.

The IP (inverse problem) is: given the set of ordered
pairs $\{f(s), h(s,t)\}$ for all $t\in [0,T]$, find $a(x)$ and $q(x)$.

We prove that IP has at most one solution by reducing
the uniqueness of the solution to IP  
to the Ramm's uniqueness theorem for the solution to elliptic boundary
value
problem [11].

This theorem says: 

Let
\begin{equation}
  Lu+\lambda u=0 \hbox { in } D, \quad u=f(s) \hbox { on }S,
  \end{equation}
and assume that the above problem is uniquely solvable for two distinct
real  values of $\lambda $. Suppose that the set of ordered pairs 
 $\{f,h\}$ is known at these values
of $\lambda$ for all $f\in H^{3/2}(S),$ 
where $h:=a(s) u_N$, and $u_N$ is the normal
derivative on $S$ of the solution to (9.5).
Then the operator $L$
is uniquely determined, that is, the functions $a(x)$ and 
$q(x)$ are uniquely determined.

We apply this theorem as follows.
 
First, we claim that the data $h(s,t)$, known for $t\in [0,T]$ are
uniquely determined for all $t>0$.
If $\delta (t)$ is replaced by a function $\eta (t)\in C^{\infty}_0(0,T),$
$\int_0^T \eta (t)dt=1$, then the data $h(s,t)$ known for $t\in [0,T]$
are uniquely determined for $t>T$.

Secondly, if this claim is established, then Laplace-transform
problem (9.1)-(9.3) to get the elliptic problem
studied in [11]:

\begin{equation}
  Lv + \lambda v=0 \hbox { in } D, \quad u=f(s) \hbox { on } S,
  \end{equation}
and the data $H(s, \lambda),$
where $v:=\int_0^{\infty} e^{-\lambda t}u(x,t)dt.$

The data $H(s,\lambda):=\int_0^{\infty} e^{-\lambda t}h(s,t)dt$
are known for all $\lambda>0$.

Thus, Ramm's theorem yields uniqueness of the determination of $L$,
and the proof is completed.

We now sketch the proof of the claim:

The solution to the time-dependent problem can be written as:

\begin{equation}
  u(x,t)=\sum_{j=0}^{\infty}e^{-\lambda_j t}c_j \phi_j(x),
  \end{equation}
where $L\phi_j(x)=\lambda_j \phi_j(x) \hbox { in } D, 
\phi_j(x)=0 \hbox { on } S, ||\phi_j(x)||_{L^2(D)}=1.$ 
The coefficients $c_j:=-\int_S f(s) a(s)\phi_{jN}(s)ds$.

Note that the series for $u(x,t)$ and the series obtained
by termwise differentiation of it with respect to $t$ converge absolutely
and uniformly in $D \times (0,\infty)$, each of the terms
is analytic with respect to $t$ in the region $\Re t>0$, and
consequently so are these series.

Therefore the functions $u(x,t)$ and $h(s,t):=u_N(s,t)$
are analytic with respect to $t$ in the region
$\Re t>0$, so the data are uniquely determined for $t>T$
as claimed.
\qed

At $t=0$ the series (9.6) is singular: it does not converge uniformly or
even in $L^2(D)$. By this reason
the above argument is formal. One can make it rigorous 
if one replaces the delta-function in (9.3) by a $C^{\infty}_0(0,T)$
function $\eta (t)$, $\int_0^T \eta (t)dt=1$, and uses the argument
similar to the one in \cite{R45}.

\end{document}